# One for Multiple: Physics-informed Synthetic Data Boosts Generalizable Deep Learning for Fast MRI Reconstruction

Zi Wang[1,20], Xiaotong Yu[1], Chengyan Wang[2], Weibo Chen[3], Jiazheng Wang[3], Ying-Hua Chu[4], Hongwei Sun[5], Rushuai Li[6], Peiyong Li[7], Fan Yang[8], Haiwei Han[8], Taishan Kang[9], Jianzhong Lin[9], Chen Yang[10], Shufu Chang[11], Zhang Shi[12], Sha Hua[13], Yan Li[14], Juan Hu[15], Liuhong Zhu[16], Jianjun Zhou[16], Meijing Lin[17], Jiefeng Guo[18], Congbo Cai[1], Zhong Chen[1], Di Guo[19], Guang Yang[20], Xiaobo Qu[1]

Magnetic resonance imaging (MRI) is a widely used radiological modality renowned for its radiation-free, comprehensive insights into the human body, facilitating medical diagnoses. However, the drawback of prolonged scan times hinders its accessibility. The k-space undersampling offers a solution, yet the resultant artifacts necessitate meticulous removal during image reconstruction. Although Deep Learning (DL) has proven effective for fast MRI image reconstruction, its broader applicability across various imaging scenarios has been constrained. Challenges include the high cost and privacy restrictions associated with acquiring large-scale, diverse training data, coupled with the inherent difficulty of addressing mismatches between training and target data in existing DL methodologies. Here, we present a novel Physics-Informed Synthetic data learning framework for Fast MRI, called PISF. PISF marks a breakthrough by enabling generalized DL for multi-scenario MRI reconstruction through a single trained model. Our approach separates the reconstruction of a 2D image into many 1D basic problems, commencing with 1D data synthesis to facilitate generalization. We demonstrate that training DL models on synthetic data, coupled with enhanced learning techniques, yields *in vivo* MRI reconstructions comparable to or surpassing those of models trained on matched realistic datasets, reducing the reliance on real-world MRI data by up to 96%. Additionally, PISF exhibits remarkable generalizability across multiple vendors and imaging centers. Its adaptability to diverse patient populations has been validated through evaluations by ten experienced medical professionals. PISF presents a feasible and cost-effective way to significantly boost the widespread adoption of DL in various fast MRI applications. Importantly, it alleviates the ethical and practical challenges associated with *in vivo* human data acquisitions, opening new horizons for advancing fast MRI technologies.

Magnetic resonance imaging (MRI) is a widely used radiological modality with radiation-free, non-invasiveness, and excellent soft-tissue contrast[1,2]. It can provide abundant and diverse information of the whole human body for medical diagnosis, through visualizing various anatomical structures, physiological functions, and tissue parameters[3-5]. However, MRI usually requires prolonged acquisition time, resulting in patient discomfort, increased risk of motion artifacts, and decreased throughput[6]. Over the past two decades, to reduce the scan time, fast MRI techniques have become widely accepted practice, through employing multi-coil parallel imaging[7,8] with k-space (i.e., raw data of MRI scanners in the Fourier domain) undersampling[9,10].

In fast MRI, a crucial step is to solve the ill-posed inverse problem of removing strong artifacts introduced by k-space undersampling, to provide high-quality and reliable images for subsequent diagnosis. Consider the imaging process as a physics-based forward operator $\mathcal{A}$, the measured k-space $\mathbf{Y}$ of an image $\mathbf{X}$ is modeled as $\mathbf{Y} = \mathcal{A}(\mathbf{X})$. Estimating a proper image $\hat{\mathbf{X}}$ from $\mathbf{Y}$ is an inverse problem. Traditionally, to solve this problem, least square and gradient descent methods[7,8,11] are applied to minimize the loss between $\mathbf{Y}$ and $\mathcal{A}(\hat{\mathbf{X}})$. To mitigate the ill-posedness, lots of hand-crafted priors are incorporated as regularization terms[9,10,12-15]. They rely on data-specific self-learning to arrive at a solution, but always suffer from the time-consuming iteration process and cumbersome parameter determination[16].

Recently, deep learning (DL) has emerged as a powerful tool for solving inverse problems in fast MRI[17-21]. Most DL methods have been demonstrated to achieve state-of-the-art results relying on large-scale and well-curated realistic training databases[22-26] that contain matched pairs of fully sampled images (i.e., ground truths) and undersampled k-space data. However, due to the complexity and universality of MRI scanning, practical applications often involve multiple contrasts, anatomies, vendors, and centers, which poses a huge challenge to DL methods. It is becoming apparent that such a conventional DL paradigm (Fig. 1) is unsuitable for multi-scenario MRI reconstruction[20], because i) realistic data acquisition is generally time-consuming, labor-intensive, and privacy-restricted. More importantly, it is impractical to collect datasets that cover all imaging scenarios, considering the dynamic nature and the diversity of MRI applications; ii) the mismatches between training and target data are inevitable practically, so the lack of robustness greatly limits the widespread usage of DL in this field. To maintain a good reconstruction performance in different scenarios, researchers pay great effort but are constrained to train multiple expert DL models specifically to handle different tasks[27-29].

[1]Department of Electronic Science, Intelligent Medical Imaging R&D Center, Fujian Provincial Key Laboratory of Plasma and Magnetic Resonance, National Institute for Data Science in Health and Medicine, Xiamen University, China. [2]Human Phenome Institute, Fudan University, China. [3]Philips Healthcare, China. [4]Siemens Healthineers Ltd., China. [5]United Imaging Research Institute of Intelligent Imaging, China. [6]Department of Nuclear Medicine, Nanjing First Hospital, China. [7]Shandong Aoxin Medical Technology Company, China. [8]Department of Radiology, The First Affiliated Hospital of Xiamen University, China. [9]Department of Radiology, Zhongshan Hospital Affiliated to Xiamen University, China. [10]Department of Neurosurgery, Zhongshan Hospital, Fudan University (Xiamen Branch), China. [11]Department of Cardiology, Shanghai Institute of Cardiovascular Diseases, Zhongshan Hospital, Fudan University, China. [12]Department of Radiology, Zhongshan Hospital, Fudan University, China. [13]Department of Cardiovascular Medicine, Heart Failure Center, Ruijin Hospital Lu Wan Branch, Shanghai Jiaotong University School of Medicine, China. [14]Department of Radiology, Ruijin Hospital, Shanghai Jiaotong University School of Medicine, China. [15]Medical Imaging Department, The First Affiliated Hospital of Kunming Medical University, China. [16]Department of Radiology, Zhongshan Hospital, Fudan University (Xiamen Branch), Fujian Province Key Clinical Specialty Construction Project (Medical Imaging Department), Xiamen Key Laboratory of Clinical Transformation of Imaging Big Data and Artificial Intelligence, China. [17]Department of Applied Marine Physics and Engineering, Xiamen University, China. [18]Department of Microelectronics and Integrated Circuit, Xiamen University, China. [19]School of Computer and Information Engineering, Xiamen University of Technology, China. [20]Department of Bioengineering, Imperial College London, United Kingdom.
Correspondence should be addressed to Xiaobo Qu (quxiaobo@xmu.edu.cn).

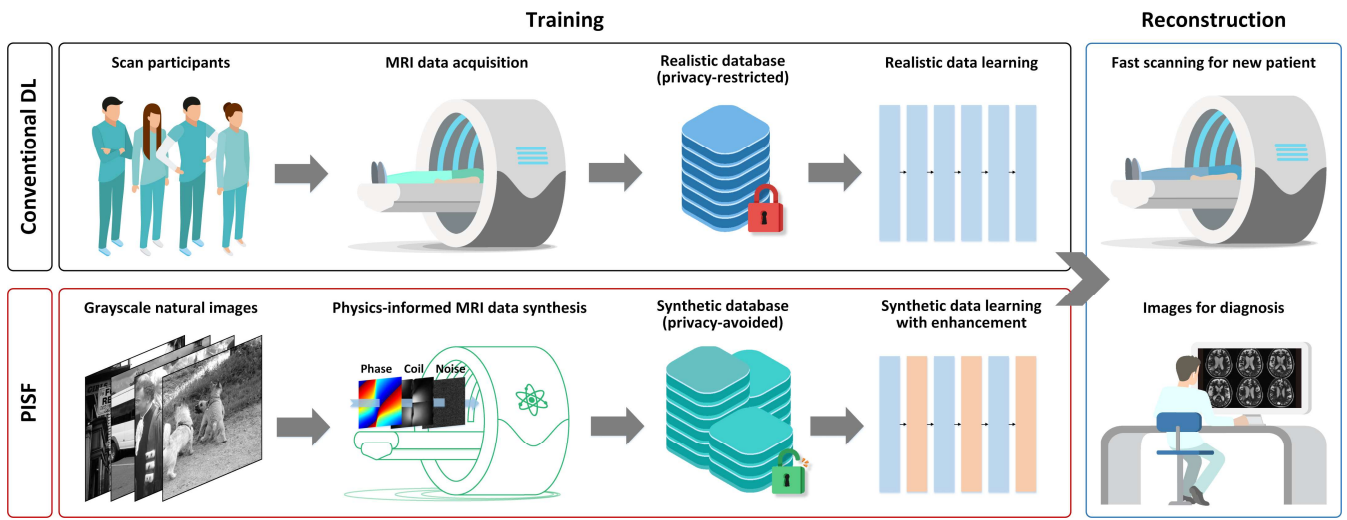

**Fig. 1 | The Overall concept of the proposed PISF.** Top: Conventional DL paradigm for fast MRI reconstruction. It heavily relies on realistic data acquisition to train DL models, which is generally costly and privacy-restricted. Bottom: The proposed PISF framework enables simplified and scaled-up data curation because numerous synthetic data are generated based on physical forward models. By integrating with enhanced learning techniques, it can perform robust *in vivo* MRI reconstruction for diagnosis. Note: Some vector images are modified from freepik.com.

Therefore, it is necessary to develop a new framework to address two significant but challenging problems in DL for fast MRI reconstruction: i) How to decrease the reliance of deep learning on large-scale realistic data? ii) How to achieve generalizable deep learning in multi-scenario imaging?

Physics-informed synthetic data learning is an emerging and promising DL paradigm for biomedical magnetic resonance[20], which is expected to address the aforementioned issues. Following plausible physical models of magnetic resonance models, such as differential equations[30,31], analytical models[32,33], or both[34,35], this paradigm can generate massive synthetic data without or with few realistic data, making learning more scalable and explainable and better protecting privacy[36]. It further introduces unique opportunities as it enables training data that may be impractical or impossible to collect in the real world[37-39]. In the MRI field, existing works strongly rely on specific acquisition protocols and anatomies to generate realistic-looking synthetic data, and then complete corresponding tasks effectively. However, these approaches lead to a decline in the universality, which makes it hard to handle multiple MRI scenarios when facing inconsistencies between partial synthetic and experimental configurations[20].

Inspired by conventional methods relying on hand-crafted priors for regularization, such as total variation[40] and wavelet transform sparsity[9,41,42], we hypothesize that DL inherently acquires a data-driven signal prior for artifacts removal through training. When realistic training data is lacking, intricate physiological effects of organs are missing, then a possible way is to search general signal priors to approximate MRI signals. We validate this concept in Supplementary Note 1, demonstrating that realistic training data is not obligatory and can be effectively substituted with appropriately generated synthetic data.

Here, we present a more versatile framework, called PISF, which uses one trained model to enable generalizable DL for multi-scenario MRI reconstruction. The overall concept of PISF is illustrated in Fig. 1. It generates large-scale and diverse synthetic data based on physical forward models for training a DL foundation network, and then utilizes the data-specific enhancement to achieve robust performance under multi-scenarios (See Supplementary Note 2 for more detailed network architecture). Remarkably, for a 2D image reconstruction problem in fast MRI, we decide to separate it into many 1D basic problems and start with the generation of 1D training signals, which is evidenced to facilitate much better generalization than 2D ones (Supplementary Note 3).

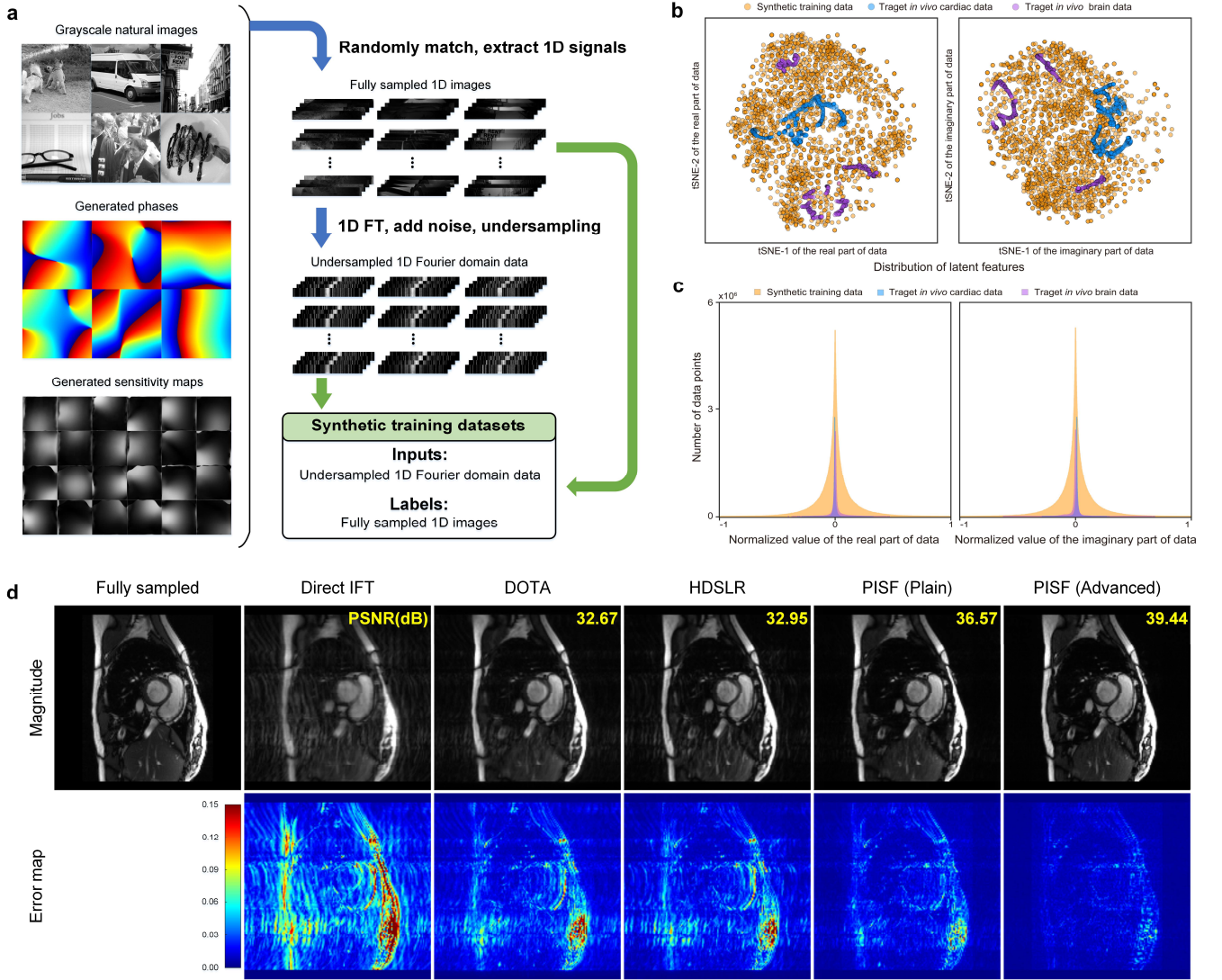

**Fig. 2 | Flowchart of the physics-informed synthetic data generation, and PISF reconstructions of the plain and advanced version. a**, following the MRI physics-based forward model, a data engine focused on generating 1D synthetic training datasets is built. **b**, the tSNE is adopted to visualize features from high dimensions. Latent feature distributions of 1D synthetic training data and 1D rows of target *in vivo* data are displayed. **c**, histograms of normalized signal intensity values for synthetic training data and target *in vivo* data, and the distribution of synthetic data has a wider intensity range than that of *in vivo* data. **d**, cardiac reconstructions of different methods under the 1D Cartesian undersampling pattern with AF=3. Note: "FT" is the Fourier transform. "Direct IFT" indicates that using only inverse Fourier transform to reconstruct undersampled k-space leads to images with strong artifacts. "DOTA" and "HDSLR" are two DL methods trained using the realistic fastMRI dataset. Because this dataset has no cardiac data, it is a mismatched reconstruction for all methods, which is fair. The mean values of PSNR are computed over all tested cases.

For data synthesis in our approach, five key components closely related to the imaging process should be considered: Magnitude $\mathbf{A}$, phase $\mathbf{P}$, coil sensitivity maps $\mathbf{S}$, noise $\mathbf{\eta}$, and undersampling operation $\mathcal{U}$. As shown in Fig. 2a, following the imaging physics-based forward model, we build a data engine focused on generating undersampled 1D synthetic Fourier domain signal $\mathbf{y}$ as:

$$\mathbf{y} = \mathcal{A}(\mathcal{P}(\mathbf{X})) = \mathcal{U}\mathcal{F}_{1D}\mathcal{P}(\mathbf{X}) + \mathbf{\eta} = \mathcal{U}\mathcal{F}_{1D}\mathbf{x} + \mathbf{\eta},$$

where $\mathcal{P}$ is the operator which randomly extracts a row from each coil data, $\mathbf{x}$ is the fully sampled 1D image, $\mathcal{F}_{1D}$ is the 1D Fourier transform (FT), $\mathbf{X} = \mathbf{A} \odot \mathbf{P} \odot \mathbf{S}$ is the fully sampled 2D image, and $\odot$ is element-wise multiplication. Specifically, we randomly i) select natural images from ImageNet[43] and convert them to grayscale as magnitude to simulate diverse image contrasts (Supplementary Note 1); ii) generate varied phases to simulate unideal experimental conditions; iii) estimate coil sensitivity maps from few realistic data in a public dataset[26] to simulate multi-coil acquisition; iv) add noise and do undersampling, to simulate different systems and imaging conditions.

Using the synthetic database, the main task of the DL network design is to narrow the gap between synthetic and realistic data. Therefore, we present the core idea of "separate-first-enhance-then" (Supplementary Fig. 3). It means that for a 2D

**Table 1 | Detailed information on datasets used for MRI reconstruction in this work.**

| Vendor/center | Index | Data type | Scan sequence | Image size | # coils | # cases | # images per case |
|---|---|---|---|---|---|---|---|
| Siemens 3.0T Prisma, Skyra, Biograph, and Tim Trio /NYU School of Medicine, USA | I | Coronal PDW knee | TSE | 320×320 | 15 | 5 HCs* | 25 |
| | II | Axial T2W brain | TSE | 320×320 | 16 | 5 HCs* | 16 |
| GE 3.0T Discovery MR 750 /Stanford University, USA | III | Sagittal PDW ankle | FSE | 384×256 | 8 | 2 HCs** | 20 |
| AOX 3.0T All-sense /Aoxin Medical, China | IV | Coronal T1W wrist | SE | 512×512 | 4 | 5 HCs | 10 |
| United Imaging 3.0T uPMR790 /United Imaging Healthcare, China | V | Axial T1-FLAIR brain | FSE-FLAIR | 224×224 | 32 | 3 HCs | 30 |
| | VI | Axial T2-FLAIR brain | FSE-FLAIR | 224×224 | 32 | 3 HCs | 30 |
| Philips 3.0T Ingenia /Zhongshan Hospital, China | VII | Axial T1W brain | TSE | 256×256 | 8 | 5 HCs | 10 |
| | VIII | Axial T2-FLAIR brain | FLAIR | 256×256 | 8 | 5 HCs | 10 |
| Siemens 3.0T Prisma /Ohio State University, USA | IX | SAX cine cardiac | TrueFISP | 160×160 | 30 | 3 HCs*** | 18 |
| | X | LAX cine cardiac | TrueFISP | 160×160 | 30 | 3 HCs*** | 18 |
| Siemens 3.0T Prisma, Skyra, Biograph, and Tim Trio /NYU School of Medicine, USA | XI | Axial T2W brain | TSE | 320×320 | 16 | 10 patients* | 16 |
| Philips 3.0T Ingenia /Zhongshan Hospital, China | XII | Axial T2-FLAIR brain | FLAIR | 256×256 | 8 | 10 patients | 10 |
| Siemens 3.0T Vida /Fudan University, China | XIII | SAX cine cardiac | TrueFISP | 246×246 | 10 | 7 patients | 12 |
| | XIV | LAX cine cardiac | TrueFISP | 224×204 | 10 | 6 patients | 12 |

Note: "HC" represents the healthy cases. "SAX" and "LAX" is the short-axis and long-axis view of cardiac MRI, respectively. *From the public dataset at https://fastmri.org. **From the public dataset at http://www.mridata.org. ***From the public dataset at https://ocmr.info. Others are in-house datasets from our collection.

reconstruction problem, we first separate it into many 1D basic problems in the training stage[28], to train a data-versatile 1D foundation network for adaptive image de-aliasing. While in the reconstruction stage, the solution is regularized by using a data-specific 2D k-space enhancement obtained from target data itself, to achieve strong generalization (Supplementary Note 2). Our proposed unrolled network follows an iteration process of a reconstruction model with image and k-space priors. It not only encourages physical behaviors like model-based iterative methods for good interpretability (Supplementary Note 4), but also integrates explicit physical operators, including data consistency[9,44], soft-thresholding[28,45], and self-consistency refinement[11,46], into the network architecture to achieve reliable reconstruction. Remarkably, although this approach is tailored towards the widely used 1D undersampling, it can naturally extend to 2D undersampling without any re-training (Supplementary Fig. 4).

Why physics-informed synthetic data learning can work? For data synthesis, although the synthetic texture may be quite different from realistic MR images, our usage of 1D learning alleviates reliance on spatial relationships and enhances the correlation between the synthetic and realistic data. For example, The latent features of the 1D synthetic data and 1D rows of target data are visualized in Fig. 2b, showing that most features of the target *in vivo* MRI data are covered by those of the synthetic training data. Besides, we can see that the statistical distribution of the synthetic data also encompasses the realistic data with a wider intensity range in the histogram (Fig. 2c). Thus, we can expect synthetic data learning to yield more versatile nonlinear mappings between images with and without artifacts, and then handle well on the realistic data.

On the other hand, the advanced DL model also plays an important role. We found that our usage of adaptive thresholding[47] and data-specific k-space enhancement[11] (Supplementary Note 2) is beneficial for better generalization. In Figs. 2d, our plain model that does not use the above two modules has already suppressed most image artifacts under synthetic data learning, and is much better than two state-of-the-art DL methods[24,25] trained using the largest publicly available but mismatched brain and knee dataset[26]. Moreover, the advanced version of PISF with the aforementioned two modules further improves peak signal-to-noise ratio (PSNR) by about 8% over the plain one, and preserves image details better. Therefore, we used the advanced version of PISF below.

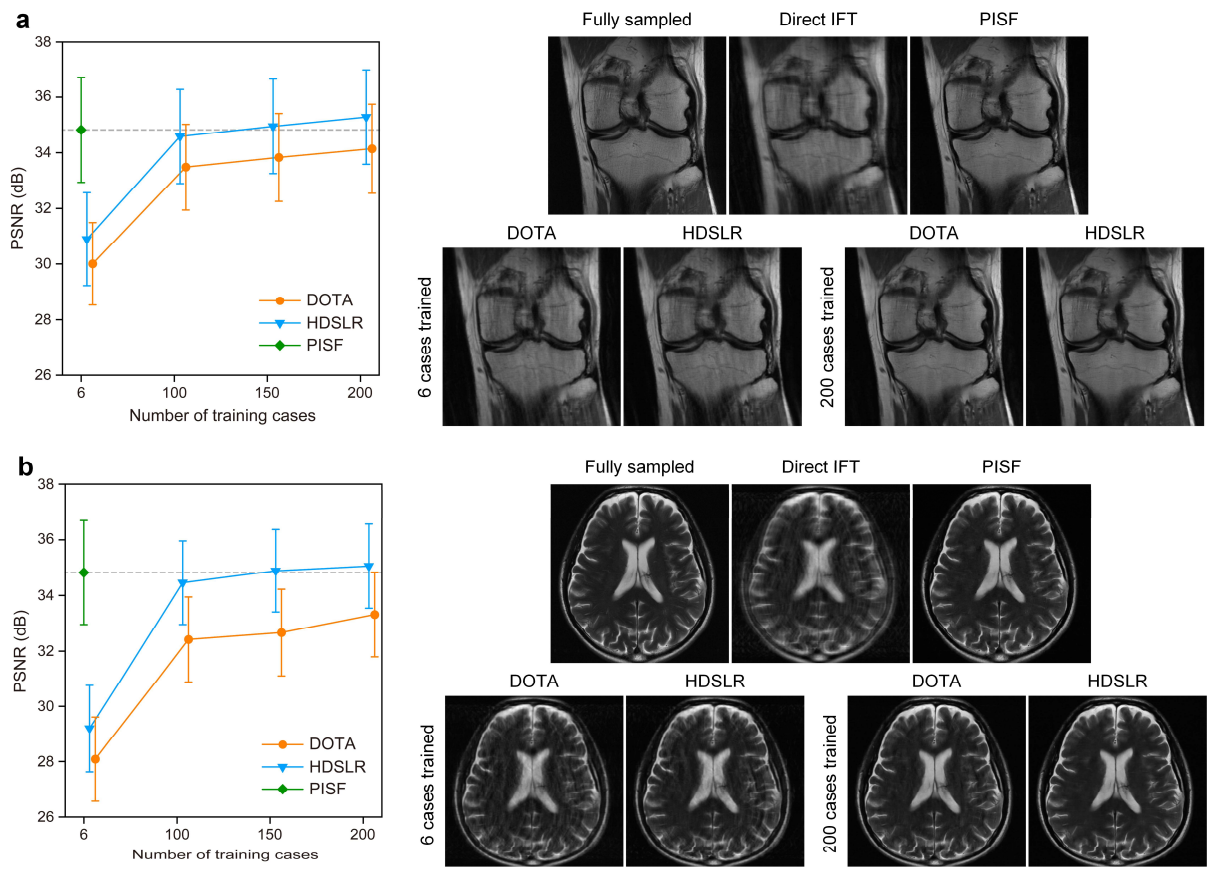

**Fig. 3 | Substitutability of PISF framework for conventional DL. a**, PDW knee reconstruction results of different methods using different number of training cases (Data I). **b**, T2W brain reconstruction results of different methods using different number of training cases (Data II). Note: PISF only estimates coil sensitivity maps from autocalibration signals of 6 cases of realistic data for synthesis, which means no fully sampled MRI data is needed. The 1D Cartesian undersampling pattern with AF=4 is used. The mean values and standard deviations of PSNR are computed over all tested cases.

In the following sections, we carefully evaluate the substitutability of the proposed PISF framework for conventional DL methods (Fig. 3)—it achieves comparable performance to realistic data training methods while reducing the demand for real-world MRI data by up to 96%. We then examine PISF's generalization ability to multi-vendor multi-center data (Fig. 4)—it can reconstruct high-quality images of 4 anatomies and 5 contrasts across 5 vendors and centers using a single trained network. More importantly, we explore the adaptability of our technique to patients (Fig. 5), where the morphology of abnormal tissues is extremely complicated and diverse, to show its reliability in clinical diagnosis—its overall image quality steps into the excellent level (i.e., 4 out of a 5-point scale) under 10 experienced doctors' evaluations.

To our knowledge, there has not yet been a such generalizable DL framework that can be widely used for fast MRI reconstruction under multiple scenarios. With a single trained model, our framework supports the reconstruction under 4 sampling patterns, 5 anatomies, 6 contrasts, 5 vendors, and 7 centers. This work demonstrates a feasible and cost-effective way to train the DL model for fast MRI reconstruction.

## Results

### Substitutability of PISF framework for conventional DL

One challenge with conventional DL methods is the dependence on large-scale realistic training data, while MRI raw data acquisition is costly or sometimes impossible. In contrast, the proposed PISF presents a novel paradigm for fast MRI. Its reconstruction of high-quality images necessitates only requires coil sensitivity information derived from a limited set of undersampled realistic data for synthesis, thereby substantially mitigating the reliance on fully sampled raw data.

To demonstrate the substitutability of the proposed method for conventional DL trained by using realistic data, we first conducted experiments (Fig. 3) on the largest publicly available dataset, fastMRI[26]. We randomly selected 200 cases (about 20 images per case) from it as the training dataset for conventional DL, including 100 knee data with 2 contrasts (PDW and PDFS) and 100 brain data with 2 contrasts (T2W and T1W), to ensure data diversity. The reconstruction test was conducted on other 5 cases of knee data and 5 cases of brain data (Data I and II in Table 1). All data were fully sampled k-space, which were first retrospectively undersampled and then used for training and testing. For the proposed method, it only estimated coil sensitivity maps from the autocalibration signals (i.e., the fully sampled region in central k-space) of 6 cases of data, to generate synthetic data (Supplementary Note 5).

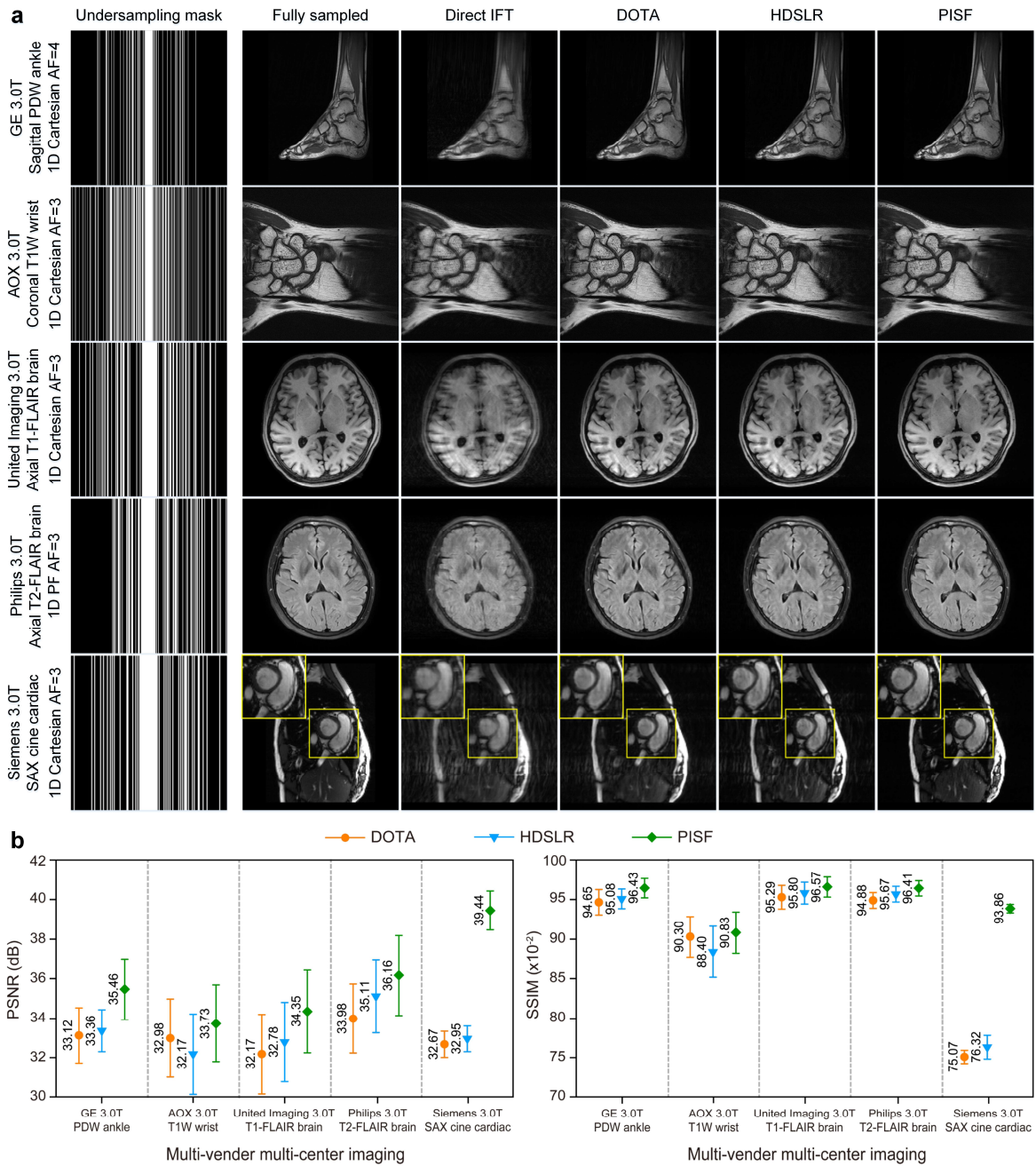

**Fig. 4 | Multi-vendor multi-center imaging. a**, multi-scenario reconstruction results using different methods. Here, they include 4 anatomies and 5 contrasts across 5 vendors and centers, under 3 undersampling scenarios (Data III, IV, V, VIII, IX). **b**, quantitative comparisons of reconstructions are shown, including PSNR and SSIM. Note: The mean values and standard deviations of PSNR and SSIM are computed over all tested cases, respectively. "1D PF" represents the 1D 3/4 partial Fourier undersampling pattern.

We compared the proposed method with two state-of-the-art DL methods: DOTA[24] and HDSLR[25], both of which were trained using large-scale realistic data. At this stage, the comparative study mainly considers matched reconstruction, which means that both training and test data are within the scope of the fastMRI dataset (Fig. 3). The training was conducted using the 1D Cartesian undersampling pattern with acceleration factor (AF) equal to 4. The AF is defined as the ratio of the number of fully sampled k-space data points to partially acquired points.

As illustrated in Fig. 3, the proposed PISF achieves high-quality image reconstruction of *in vivo* knee and brain data, both visually and quantitatively. Notably, this accomplishment is attained by utilizing sensitivity information derived exclusively from solely 6 cases of realistic data. For further insights into the impact of varying this quantity, refer to Supplementary Note 5. As for compared DOTA and HDSLR, their networks are not sufficiently trained with 6 cases, resulting in obvious image artifacts and lost details. Only when the number of training cases significantly increases, the satisfied reconstruction quality can be obtained. Specifically, DOTA can reconstruct good images using 200 training cases, but it is still inferior to HDSLR and the proposed PISF, which may be due to its relatively simple end-to-end network architecture. HDSLR, with its advanced model-based unrolled network architecture, can use 150 training cases to achieve comparable results to PISF and further improve PSNR by 0.46 and 0.23 dB for knee and brain reconstructions, respectively, in 200 cases.

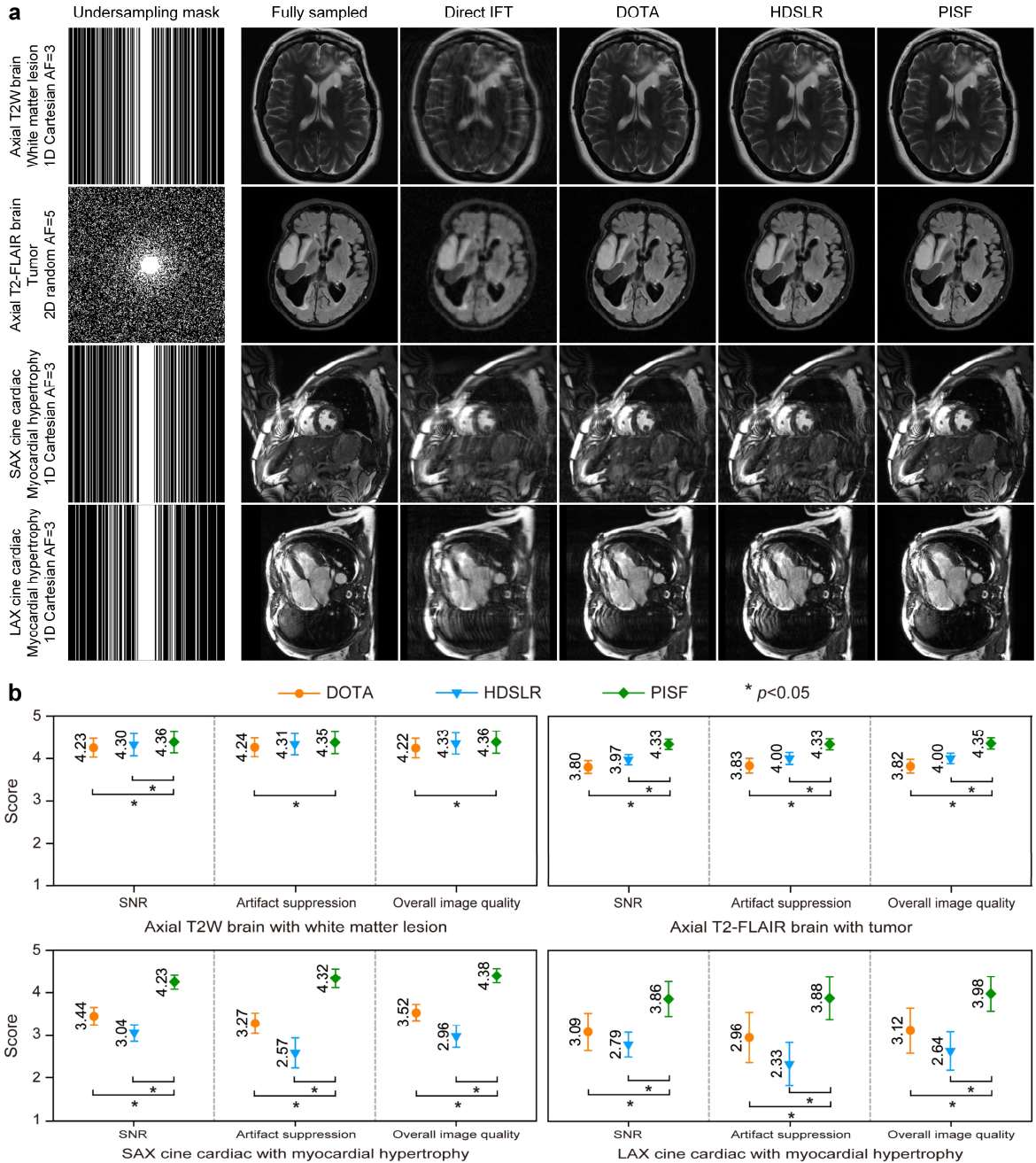

**Fig. 5 | Adaptability to patients with pathologies. a**, different types of patients' reconstruction results using different methods (Data XI, XII, XIII, XIV). **b**, score comparisons of the reader study are shown. Note: The mean values and standard deviations are computed over all tested patients, respectively. $p<0.05$ indicates the differences between the compared methods are statistically significant.

We observed that the proposed PISF achieves performance comparable to the advanced DL method HDSLR, while reducing the demand for MRI raw data by up to 96% (from 150 to 6 cases).

Thus, one of the main contributions of this work is that it can be efficiently applied to *in vivo* MRI reconstruction by using physics-informed synthetic data learning. Its performance approaches or even surpasses conventional DL, especially as an excellent substitute when realistic data availability is severely low.

**Multi-vendor multi-center imaging**

Multi-vendor multi-center data harmonization is an important issue for healthcare[48,49]. Due to the diversity of MRI, these data can show substantial heterogeneity in the image distribution (e.g., due to multiple contrasts, anatomies, and scanners) or the imaging operator (e.g., due to multiple image sizes, undersampling patterns, and AFs). It imposes forward higher demands on the generalizability of DL methods.

To examine the proposed PISF's generalization ability in multi-scenario MRI reconstruction, we conducted experiments on multi-vendor multi-center imaging (Data III-X in Table 1), including 4 anatomies and 5 contrasts across 5 vendors and centers, under varied undersampling scenarios. In the conducted experiments, all DL methods utilized the pre-trained networks from the preceding section without undergoing additional re-training or fine-tuning. This strategy was adopted due to the impracticality of allocating considerable resources to repeatedly adjust networks to align with the target settings.

As shown in Fig. 4a and Supplementary Note 6, our method presents robust and nice performance in multi-scenario imaging, while other DL methods yield unsatisfactory results or even fail to remove image artifacts. Specifically, DOTA and HDSLR provide the images with poor artifacts suppression, since they are hard to handle mismatches between their training data and these multi-scenario data. When reconstructing the completely unseen anatomical structure, such as cardiac MRI, both DOTA and HDSLR completely fail and leave strong image artifacts and blurring (Their PSNRs are about 17% lower than our PISF). On the contrary, the proposed PISF works well in artifacts suppression and structural preservation among *in vivo* ankle, wrist, brain, and cardiac data with different contrasts, which can also be verified in the comparative analysis of the quantitative evaluation (Fig. 4b and Supplementary Note 6). Moreover, the consistently superior performance of PISF under matched undersampling settings and more insights can be found in Supplementary Note 7.

These results imply that, our PISF framework is proposed as a robust and versatile technique that enables generalizable DL in a wide range of fast MRI scenarios without further re-training.

**Adaptability to patients with pathologies**

Considering the morphology of pathological tissues is extremely complicated and diverse, we further examine the adaptability of our method to patients, which is essential in clinical diagnosis.

We reconstructed 4 types of patient data (Data XI-XIV in Table 1), including T2W brain with non-specific white matter lesion (10 cases), T2-FLAIR brain with tumor (10 cases), and cine cardiac with myocardial hypertrophy (7 cases of SAX and 6 cases of LAX). In these experiments, all DL methods also used the same trained models as before. Specifically, considering that patient datasets are very limited and the quantity is hard to meet training requirements, we still use the healthy-case-trained models of compared methods to reconstruct patients directly. Correspondingly, our PISF adheres to the same practice, refraining from incorporating any pathological information into the data synthesis. It ensures a fair and unbiased comparison, as all methods including PISF, have not been exposed to any pathological tissues.

We invited 5 neuro readers (4 radiologists with 6/13/20/30 years' experience and 1 neurosurgeon with 9 years' experience) and 5 cardiac readers (3 radiologists with 3/8/12 years' experience and 2 cardiologists with 12/13 years' experience), to independently evaluate the brain or cardiac reconstructed images from a diagnostic perspective. They were blind to all patient information and reconstruction methods, while fully sampled images were provided as the reference. Three clinical concerned criteria including signal-to-noise ratio (SNR), artifacts suppression, and overall image quality were used in the reader study. The score of each criterion had a range from 0 to 5 with a precision of 0.1 (i.e., 0~1: Non-diagnostic; 1~2: Poor; 2~3: Adequate; 3~4: Good; 4~5: Excellent). The difference in scores between PISF and compared methods was analyzed using the Wilcoxon signed-rank test, and $p<0.05$ was considered statistically significant.

Fig. 5 shows that, the proposed PISF scores are higher than or very close to 4 on all patient data in all three criteria. From a diagnostic perspective, its image quality steps into an excellent level and is significantly superior to the two compared methods DOTA and HDSLR. It should be noted that, DOTA and HDSLR have strong image artifacts in the reconstructed images of cardiac patient data, resulting in low scores that can only reach the adequate level. Thus, both methods need to be used with caution if a new training is not applied. The full report for all individual readers is presented in Supplementary Note 8.

This section demonstrates that, our technique is suitable for fast MRI of patients and provides reliable image reconstructions. From the perspective of experienced doctors, our PISF surpasses the realistic-healthy-case-trained methods and its overall image quality steps into the excellent level for clinical diagnosis.

**Discussion and conclusion**

In this study, we have showcased the efficacy of our physics-informed synthetic data learning in endowing reliability and adaptability to the DL model for addressing the inverse problem inherent in fast MRI reconstruction across diverse scenarios. It is imperative to clarify that the primary emphasis of this research does not lie in the direct comparison of our method's performance against existing DL methods under conditions where a large amount of task-domain data is available. The rationale for this work is rooted in the crucial and pressing need within clinical MRI applications to reliably reconstruct unseen data, a critical aspect that transcends the comparative evaluation under data-rich scenarios.

Our experiments on varied tasks show that the performance of our synthetic-data-trained PISF meets or exceeds the models trained on a matched realistic dataset, while PISF further owns remarkable generalizability in multi-vendor multi-center imaging and patients' diagnosis. It means that, compared with acquiring numerous MRI data from humans, generating synthetic data is a viable and cost-effective resource for developing DL models.

Thanks to the flexibility of our PISF framework in data synthesis and separable learning, we anticipate that it is also applicable to higher dimensional imaging, such as dynamic, diffusion, and quantitative MRI, by considering organ motion, diffusion information, and tissue parameters into physics model, while adjusting network design accordingly. In addition, optimizing/replacing components in data synthesis (e.g., using easy-to-collect DICOM images as the magnitude for the specific task), integrating emerging network architecture[50,51], transfer learning[29] (e.g., taking our PISF as a pre-trained model in Supplementary Note 9), and using synthetic data to enrich the diversity of realistic data (Supplementary Note 10) are promising directions, which can provide continuous vitality for the physics-informed synthetic data learning in the face of increasingly diverse MRI applications.

Limitations of the current PISF are like other methods used for fast MRI: i) Unsatisfactory reconstructions under high ill-posedness, e.g., 1D Cartesian undersampling with extremely high acceleration (AF=8). Even though, it demonstrates somewhat more robust results than other state-of-the-art DL methods (Supplementary Note 11). ii) Care should be taken when applying the proposed method to more imaging scenarios that are not involved in this work, as our experiments cannot cover all MRI scans. For example, when facing other scan sequences or anatomies, testing should be conducted and proper modifications may be necessary. Besides, there are several aspects worth investigating in future: i) Prospective undersampling. The proposed PISF has been successfully applied to various retrospective undersampling scenarios, and we believe it can also be used in prospective undersampling as long as it has the autocalibration signals for our k-space enhancement. ii) More demanding and challenging applications, such as 3D and non-Cartesian imaging. For 3D imaging, the segmentation of 3D images into a series of 2D images for independent processing is feasible, given that each 2D image typically shares a same 2D undersampling pattern. Given PISF's demonstrated proficiency in reconstructing 2D undersampled data, it holds promise for handling 3D datasets[52]. Concerning non-Cartesian data, while PISF may not be directly applicable, a plausible approach involves first gridding the data into Cartesian coordinates[53] and subsequently proceeding with reconstruction[54,55].

In summary, we present a proof-of-concept demonstration of applying physics-informed synthetic data learning to reconstruct high-quality MRI images of different contrasts, anatomies, vendors, centers, and pathologies from fast sampled data. It opens a new avenue for the widespread application of DL in MRI, without the need to consider the intractable ethical and practical issues of *in vivo* human data acquisitions.

## Methods
### Data engine for physics-informed data synthesis
According to the aforementioned imaging physics-based forward model, our data engine generates paired 1D synthetic signals as follows: 1) Randomly select 4000 2D natural images from ImageNet[43] and convert them to grayscale as magnitude to simulate diverse image contrast. 2) Randomly generate phase through Fourier truncation with size 2~5 and amplitude normalization of the complex white Gaussian noise[14]. 3) Randomly select coil sensitivity maps estimated from 6 cases of realistic data in fastMRI dataset[26] using ESPIRiT[56]. 4) Randomly match the above three components and extract 1D images with size 320. 5) Perform the 1D FT, add complex white Gaussian noise within a range of SNR from 10 to 80 dB randomly, and use 1D Cartesian undersampling with AF=4 sequentially to obtain an undersampled 1D Fourier domain signal. Notably, to enrich the diversity of the datasets, with the same AF, the undersampling mask is different for each data. The AF is defined as the ratio of the number of fully sampled k-space data points to that of partially acquired ones.

Thus, 640000 paired 1D synthetic data $(\mathbf{y}^t, \mathbf{x}^{ref,t})$ are generated for network training, where $\mathbf{x}^{ref}$ is the label 1D image, $t = 1, 2, ..., T$ denotes the $t^{th}$ training sample. Among them, 90% (576000 pairs) is for training to obtain the optimal network parameters, and 10% (64000 pairs) is for validation to adjust the hyperparameters and evaluate the reconstruction performance of the network preliminarily.

### Implementation of the PISF framework
The proposed method is an unrolled network that follows an iteration process of a reconstruction model with image and k-space priors, and also integrates explicit physical operators, including data consistency[9,44], soft-thresholding[28,45], and self-consistency refinement[11,46], to achieve reliable reconstruction (Supplementary Fig. 3). The total number of network phase $K=10$ is an optimal trade-off between the reconstruction performance and time consumption. At each network phase, the image de-aliasing module consists of 2 convolutional layers and 4 ResBlocks[57]. Each convolutional layer contains 64 1D convolution filters of size 3, followed by batch normalization (BN) and a ReLU as the activation function. The adaptive threshold module[47] has 2 convolutional layers, each containing 64 1D convolution filters of size 1. Besides, the 5×5 SPIRiT kernel[11] is used for k-space enhancement. Additionally, although this approach is tailored towards the widely used 1D undersampling, it is natural to generalize it to handle 2D undersampling by separately processing rows and columns of the image in the trained de-aliasing modules. See Supplementary Note 2 for more details on the DL model.

In the training stage, the data-specific k-space enhancement is not involved. The 1D foundation network is trained for 100 epochs with the Adam optimizer[58]. Its initial learning rate is set to

0.001 with an exponential decay of 0.99, while the batch size is 128. The loss function is defined as:

$$\mathcal{L}(\mathbf{\Theta}) = \frac{1}{KT}\sum_{k=1}^{K}\sum_{t=1}^{T}\left\|\mathbf{x}^{ref,t} - \mathbf{x}^{(k),t}\right\|_{2}^{2},$$

where $T$ is the number of training samples, and $\mathbf{x}^{ref,t}$ is the label of the $t^{\text{th}}$ training sample.

In the reconstruction stage, for undersampled k-space data, we can reconstruct them through the trained 1D foundation network with data-specific 2D k-space enhancement. Notably, a single pre-trained PISF model can be effectively applied to image reconstructions in multi-scenario fast MRI.

The proposed PISF framework was implemented on a server equipped with dual Intel Xeon Silver 4210 CPUs, 256 GB RAM, and the Nvidia Tesla T4 GPU (16 GB memory) in PyTorch 1.10. The typical training took about 23 hours, and the typical reconstruction time per image was about 0.17 seconds.

**MRI data preparation**

The *in vivo* MRI data used in this work included publicly available datasets[26,59] and in-house datasets from our collection. Their study protocol was Institutional Review Board approved and informed consent was obtained from volunteers before examination. The detailed information for all datasets is provided below.

For training conventional DL methods, 200 cases (about 20 images per case) with 4 contrasts from fastMRI dataset[26] were used, to ensure data diversity. Specifically, there were 100 coronal proton-density-weighted knee without (PDW, 50 cases) and with fat suppression (PDFS, 50 cases), and 100 axial brain with T2-weighted (T2W, 50 cases) and T1-weighted (T1W, 50 cases). Besides, other 5 cases of PDW knee and 5 cases of T2W brain data were used for testing. Notably, images from this dataset were center-cropped into size 320×320.

To examine the generalizability of DL methods in multi-scenario MRI reconstruction, the multi-vendor multi-center datasets covered 4 anatomies (Ankle, wrist, brain, and cardiac MRI) and 5 contrasts (PDW, T1W, T1-FLAIR, T2-FLAIR, and cine MRI) across 5 vendors (GE, AOX, United Imaging, Philips, and Siemens) and centers. Notably, image sizes from these datasets were different, and we decided not to process them to the same size to simulate clinical practice. More details about them are listed in Table 1.

The patient dataset included 20 cases of axial brain data, 7 cases of short-axis view cardiac data, and 6 cases of long-axis view cardiac data in total (Table 1). Specifically, there were 10 T2W brain data with non-specific white matter lesion from fastMRI dataset, 10 T2-FLAIR brain data with tumor, 7 short-axis view and 6 long-axis view cine cardiac data with myocardial hypertrophy from our collection.

All datasets were fully sampled k-space data, and then undersampling retrospectively according to different patterns and AFs. The existence of fully sampled data would be helpful serving as the ground truth for objective evaluation and the reference for reader study. To reduce computational complexity, coil compression[60] was used to keep the number of coils at 8 in all datasets.

**Evaluation criteria**

To quantitatively evaluate the reconstruction performance, we utilized 2 objective and 3 subjective evaluation criteria.

Two objective criteria include the peak signal-to-noise ratio (PSNR) and structural similarity index (SSIM)[61]. Specifically, a higher PSNR (or a higher SSIM) indicates the less image distortions (or better image details preservation).

Since objective criteria might not reflect image quality in terms of diagnostically important features, three clinical-concerned subjective criteria including SNR, artifacts suppression, and overall image quality were used for the reader study. The fully sampled images were provided as the reference. The score of each criterion had a range from 0 to 5 with precision of 0.1 (i.e., 0~1: Non-diagnostic; 1~2: Poor; 2~3: Adequate; 3~4: Good; 4~5: Excellent). The reader study was conducted online through a cloud platform (https://csrc.xmu.edu.cn/CloudBrain.html), called CloudBrain-ReconAI[62,63].

**Compared methods**

We compared the proposed method with two state-of-the-art DL methods: DOTA[24] and HDSLR[25], both of which were trained using large-scale realistic data from the largest publicly available dataset[26]. DOTA and HDSLR were selected since they conducted dual-domain (Fourier and image domains) learning and this strategy can greatly improve the reconstruction performance. Differently, DOTA owns a relatively simple end-to-end network architecture, while HDSLR is an unrolled network inspired by a model with joint k-space and image-domain priors. They were implemented according to shared codes with typical settings.

## Data availability

The demo data and the trained PISF model for testing will be shared at
https://github.com/wangziblake/PISF
https://csrc.xmu.edu.cn/CloudBrain.html
All used public datasets are available on their websites, including https://fastmri.org, http://www.mridata.org, and https://ocmr.info. Other in-house MRI datasets from our collection are available from the corresponding author upon reasonable request.

## Acknowledgements

The authors thank Yewei Chen for assisting in cardiac MRI data processing and helpful discussions; Drs. Michael Lustig, Dosik Hwang, and Mathews Jacob for sharing their codes online. This work was supported in part by the National Natural Science Foundation of China under grants 62331021, 62122064, and 62371410, Natural Science Foundation of Fujian Province of China under grants 2023J02005, 2021J011184, and 2022J011425, President Fund of Xiamen University under grant 20720220063, UKRI Future Leaders Fellowship under grant MR/V023799/1, Xiamen University Nanqiang Outstanding Talents Program, and China Scholarship Council under grant 202306310177.

## Author contributions

X. Qu and Z. Wang conceived the idea and designed the study, X. Qu supervised the project. Z. Wang and X. Yu implemented the method and produced the results. Z. Wang drew all the figures and tables for the manuscript and supplementary. C. Wang, W. Chen, J. Wang, Y.-H. Chu, H. Sun, R. Li, L. Zhu, J. Zhou, and P. Li helped to acquire and collect in-house brain, cardiac, and wrist MRI datasets. F. Yang, H. Han, T. Kang, J. Lin, and C. Yang conducted the reader study on patients' brain data. S. Chang, Z. Shi, S. Hua, Y. Li, and J. Hu conducted the reader study on patients' cardiac data. The manuscript was drafted by Z. Wang and improved by Z. Wang, M. Lin, J. Guo, C. Cai, Z. Chen, D. Guo, G. Yang, and X. Qu. X. Qu and D. Guo acquired research funds and provided all the needed resources.

## Competing interests

The authors declare no competing interests.

# Supplementary Information

# One for Multiple: Physics-informed Synthetic Data Boosts Generalizable Deep Learning for Fast MRI Reconstruction

Zi Wang[1,20], Xiaotong Yu[1], Chengyan Wang[2], Weibo Chen[3], Jiazheng Wang[3], Ying-Hua Chu[4], Hongwei Sun[5], Rushuai Li[6], Peiyong Li[7], Fan Yang[8], Haiwei Han[8], Taishan Kang[9], Jianzhong Lin[9], Chen Yang[10], Shufu Chang[11], Zhang Shi[12], Sha Hua[13], Yan Li[14], Juan Hu[15], Liuhong Zhu[16], Jianjun Zhou[16], Meijing Lin[17], Jiefeng Guo[18], Congbo Cai[1], Zhong Chen[1], Di Guo[19], Guang Yang[20], Xiaobo Qu[1]

1. Department of Electronic Science, Intelligent Medical Imaging R&D Center, Fujian Provincial Key Laboratory of Plasma and Magnetic Resonance, National Institute for Data Science in Health and Medicine, Xiamen University, China.
2. Human Phenome Institute, Fudan University, China.
3. Philips Healthcare, China.
4. Siemens Healthineers Ltd., China.
5. United Imaging Research Institute of Intelligent Imaging, China.
6. Department of Nuclear Medicine, Nanjing First Hospital, China.
7. Shandong Aoxin Medical Technology Company, China.
8. Department of Radiology, The First Affiliated Hospital of Xiamen University, China.
9. Department of Radiology, Zhongshan Hospital Affiliated to Xiamen University, China.
10. Department of Neurosurgery, Zhongshan Hospital, Fudan University (Xiamen Branch), China.
11. Department of Cardiology, Shanghai Institute of Cardiovascular Diseases, Zhongshan Hospital, Fudan University, China.
12. Department of Radiology, Zhongshan Hospital, Fudan University, China.
13. Department of Cardiovascular Medicine, Heart Failure Center, Ruijin Hospital Lu Wan Branch, Shanghai Jiaotong University School of Medicine, China.
14. Department of Radiology, Ruijin Hospital, Shanghai Jiaotong University School of Medicine, China.
15. Medical Imaging Department, The First Affiliated Hospital of Kunming Medical University, China.
16. Department of Radiology, Zhongshan Hospital, Fudan University (Xiamen Branch), Fujian Province Key Clinical Specialty Construction Project (Medical Imaging Department), Xiamen Key Laboratory of Clinical Transformation of Imaging Big Data and Artificial Intelligence, China.
17. Department of Applied Marine Physics and Engineering, Xiamen University, China.
18. Department of Microelectronics and Integrated Circuit, Xiamen University, China.
19. School of Computer and Information Engineering, Xiamen University of Technology, China.
20. Department of Bioengineering, Imperial College London, United Kingdom.

Correspondence should be addressed to Xiaobo Qu (quxiaobo@xmu.edu.cn).

# Supplementary Note 1. Insights on Training Data and Network Learning

To solve the inverse problem in fast MRI, conventional methods rely on hand-crafted priors for regularization, such as total variation[1] and wavelet transform sparsity[2-4], to achieve image reconstruction. Inspired by them, we hypothesize that deep learning inherently acquires a data-driven signal prior for artifacts removal through training. When realistic training data is lacking, intricate physiological effects of organs are missing, then a possible way is to search general signal priors to approximate MRI signals. Therefore, realistic training data may not be obligatory and could be effectively substituted with appropriately generated synthetic data.

To verify this idea, we train our network using the synthetic signal priors of total variation (piecewise constant features in Supplementary Fig. 1(a)), Daubechies wavelet (piecewise smooth features in Supplementary Fig. 1(b)), and compare their latent features (visualized using tSNE in Supplementary Figs. 1(e) and (f)) to those of the proposed synthetic data with grayscale natural images (Supplementary Fig. 1(c)) and the realistic fastMRI dataset[5] (including knee and brain).

All the synthetic data can cover most features of realistic MRI data (Supplementary Figs. 1(e)-(g)), thus the realistic image could be approximately reconstructed well (Supplementary Fig. 2). However, the synthetic piecewise constant or piecewise smooth features lead to staircase or blurring artifacts (the 2$^{nd}$ and 3$^{rd}$ columns of Supplementary Fig. 2(a)), respectively, as these features are simplistic and lack of diverse. On the contrary, the proposed synthetic data starts from grayscale natural images with more sophisticated features, providing improved reconstructions and achieving sub-optimal results comparable to the realistic knee data training (the 5$^{th}$ and 6$^{th}$ columns of Supplementary Fig. 2(a)).

Another example shows the limitation of lacking target-domain realistic training data. Assuming that cardiac images are the reconstruction target but the cardiac training data is not available, one may turn to use easy-to-access knee and brain images in fastMRI dataset[5], which may miss some latent features for cardiac reconstructions (Supplementary Fig. 1(h)). In this case, the synthetic signal prior from grayscale natural images is more versatile and has strong generalizability (The tSNE visualization in Supplementary Fig. 1(g) also supports this view), resulting in better reconstruction than realistic-data-trained one (the 5$^{th}$ and 6$^{th}$ columns of Supplementary Fig. 2(b)).

These results demonstrate our idea about the essence of network training is reasonable, and also indicates that well-designed synthetic data can achieve comparable results with realistic data and better generalization on unseen data.

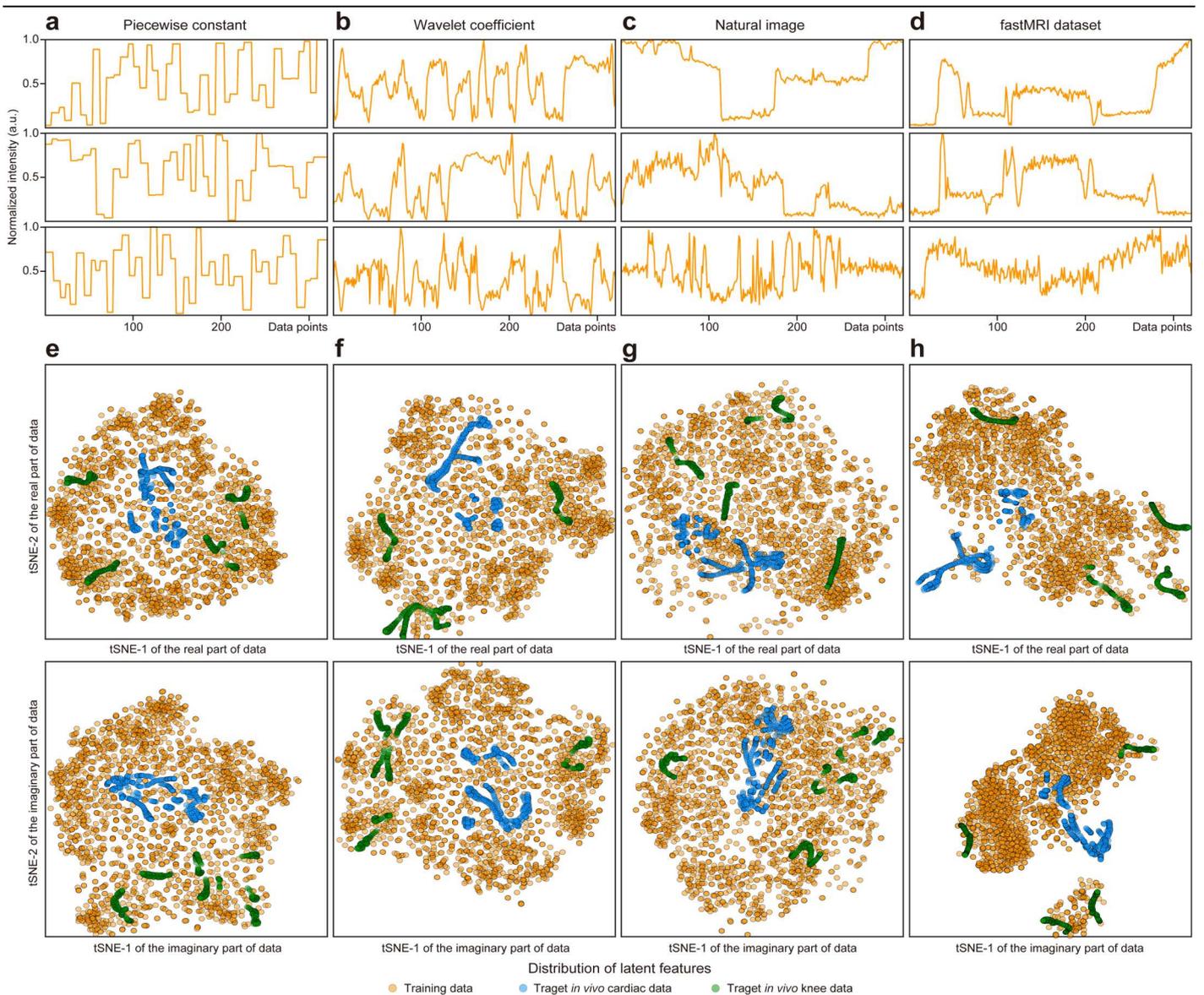

**Supplementary Fig. 1 | Visualization of different types of training data. a-d**, typical different 1D training signals. Here, we show three magnitude signals of each types of data. **e-h**, tSNE visualizations of the corresponding top data. Latent feature distributions of 1D training data and 1D rows of target *in vivo* data are displayed. The tSNE is adopted to visualize features from high dimension. Note: The data scale (size and quantity) of different types of training data is same. We then use these data to independently train four corresponding models (based on the same network architecture and training setting) for subsequent reconstructions.

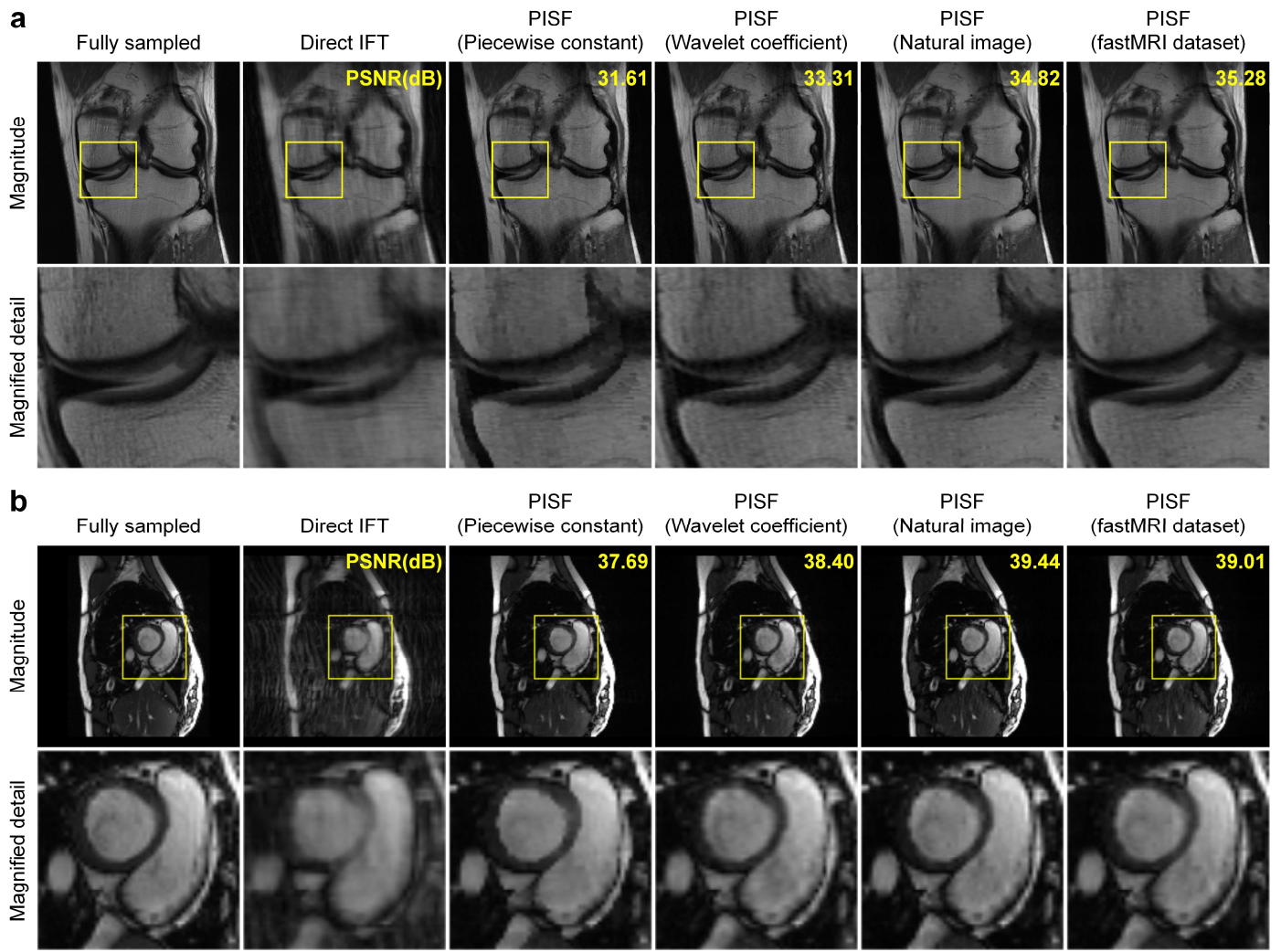

**Supplementary Fig. 2 | Comparison results using different types of training data. a**, knee reconstruction results under the 1D Cartesian undersampling pattern with AF=4 (Data I). **b**, cardiac reconstruction results under the 1D Cartesian undersampling pattern with AF=3 (Data IX). Note: The mean values of PSNR are computed over all tested cases.

# Supplementary Note 2. Methodology

In this section, we first introduce the 1D foundation network, which involves several alternating data-versatile image de-aliasing modules and data consistency modules. Then, to make use of the information of target data during reconstruction, the data-specific k-space enhancement module is integrated (Supplementary Fig. 3). Besides, an extension scheme for 2D undersampling is provided.

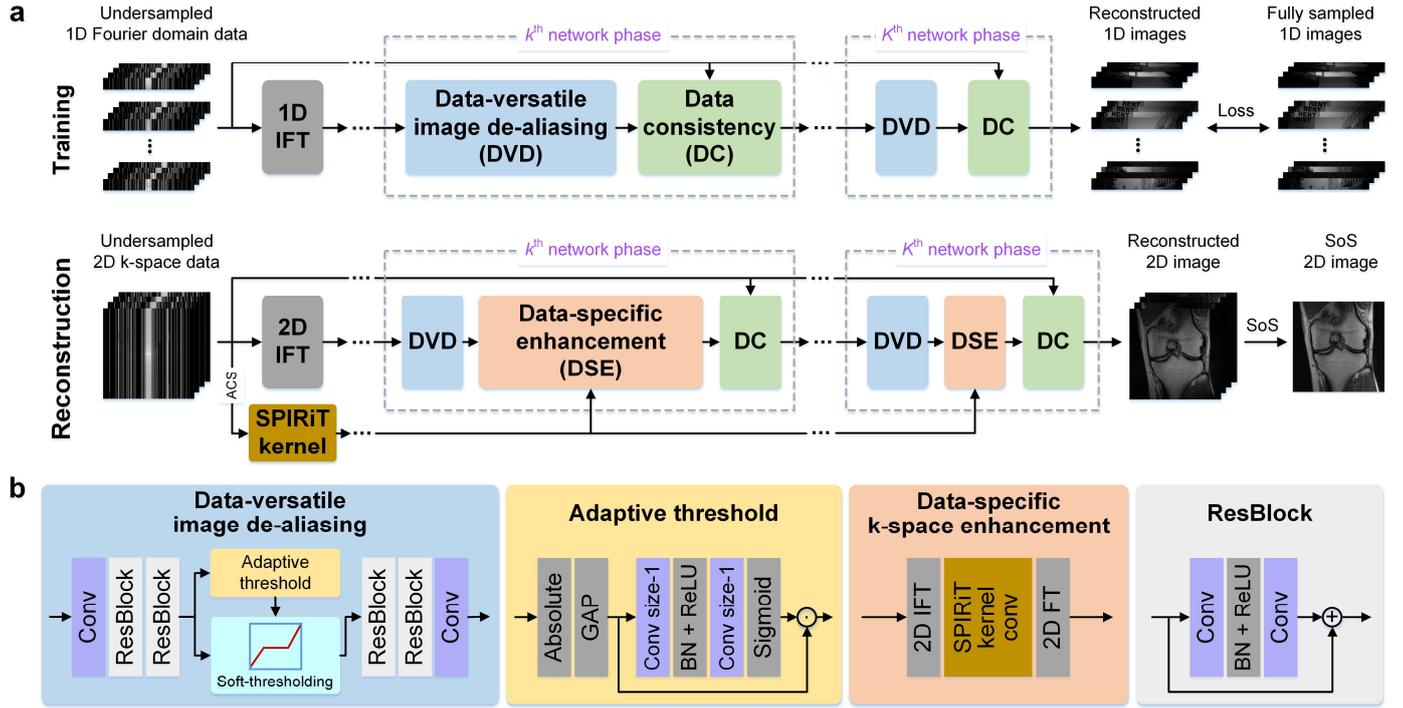

**Supplementary Fig. 3 | The network architecture of PISF for multi-scenario fast MRI reconstruction. a**, the recursive network architecture in training and reconstruction stages. With the increase of the network phase, artifacts are gradually removed, and a high-quality reconstructed image can be obtained finally. **b**, detailed structures of the network modules. Note: "Conv" is the convolution, "BN" is the batch normalization, and "GAP" is the global average pooling. "SoS" means that the reconstructed multi-coil images are finally displayed after combining by the square root of sum of squares.

## 2.1 Foundation reconstruction network

Here, we first formulate the 1D reconstruction model of the 1D image $\mathbf{x}$ with the learned deep prior:

$$\min_{\mathbf{x}} \left\| \mathbf{y} - \mathcal{U}\mathcal{F}_{1D}\mathbf{x} \right\|_2^2 + \lambda \left\| \mathbf{x} - \mathcal{D}(\mathbf{x}) \right\|_2^2, \qquad (1)$$

where $\mathcal{D}$ is the learned image de-aliasing module, $\lambda$ is the regularization parameter. The (1) can be solved by alternating two sub-problems[6], and the $k^{th}$ iteration is:

$$\begin{cases} \mathbf{d}^{(k)} = \mathcal{D}(\mathbf{x}^{(k-1)}) \\ \mathbf{x}^{(k)} = \arg\min_{\mathbf{x}} \|\mathbf{y} - \mathcal{U}\mathcal{F}_{1D}\mathbf{x}\|_2^2 + \lambda \|\mathbf{x} - \mathbf{d}^{(k)}\|_2^2 \\ \qquad = (\mathcal{F}_{1D}^* \mathcal{U}^* \mathcal{U} \mathcal{F}_{1D} + \lambda)^{-1}(\mathcal{F}_{1D}^* \mathcal{U}^* \mathbf{y} + \lambda \mathbf{d}^{(k)}) \end{cases}, \qquad (2)$$

where the superscript * represents the adjoint operation, and $\mathcal{F}_{1D}^*$ is the 1D inverse Fourier transform (IFT).

Once the overall number of iterations $K$ is fixed, the iteration process in (2) can be viewed as an unrolled deep network with $K$ phase. Each network phase consists of two modules: The image de-aliasing module and data consistency module, which correspond to the first and second step of (2), respectively. We rely on end-to-end training using the aforementioned synthetic datasets to learn the weights in $\mathcal{D}$ and set $\lambda$ as a trainable parameter. If the regularization can yield improved reconstructions, high values of $\lambda$ would be learned during the training process.

**2.1.1 Data-versatile image de-aliasing module**

Our network is trained using large-scale and diverse synthetic datasets, and then applied to the realistic MRI data reconstruction. Therefore, narrowing the gap between synthetic and realistic data is a key task of the network module design.

Recently, under the principle of meta-learning[7], an adaptive thresholding network has been proposed[8], which introduces a subnetwork to adjust artifact-removal-related thresholds according to the input. It repairs the knowledge missed in synthetic data learning and adapts well to realistic biological spectrum reconstruction[9].

Inspired by the success in the previous work[8], we attempt to design a module with adaptive thresholding, for data-versatile image artifacts removal. The overall module is designed as:

$$\mathbf{d}^{(k)} = \mathcal{D}(\mathbf{x}^{(k-1)}) = \mathcal{N}_2[soft(\mathcal{N}_1(\mathbf{x}^{(k-1)}); \boldsymbol{\theta}^{(k)})], \qquad (3)$$

where $soft(x; \rho) = \max\{|x| - \rho\} \cdot x/|x|$ is the element-wise soft-thresholding, $\boldsymbol{\theta}^{(k)}$ is the threshold, $\mathcal{N}_1$ and $\mathcal{N}_2$ are two multi-layer 1D CNNs. To be more intuitive, we decompose it into several subnetworks:

$$\begin{cases} \mathbf{r}^{(k)} = \mathcal{N}_1(\mathbf{x}^{(k-1)}) \\ \boldsymbol{\theta}^{(k)} = \mathcal{A}(\mathbf{r}^{(k)}) = \boldsymbol{\alpha}^{(k)} \odot \mathbf{g}^{(k)} \\ \mathbf{d}^{(k)} = \mathcal{N}_2[soft(\mathbf{r}^{(k)}; \boldsymbol{\theta}^{(k)})] \end{cases}, \qquad (4)$$

where $\mathcal{A}$ is a subnetwork used to determine the threshold adaptively. After passing a multi-layer 1D CNN

$\mathcal{N}_1$, a global average pooling (GAP) is applied to the absolute values of $\mathbf{r}^{(k)}$ to obtain global features $\mathbf{g}^{(k)}$. Then, $\mathbf{g}^{(k)}$ is fed into a two-layer 1D CNN which contains Sigmoid function at the end, so that the scale value $\boldsymbol{\alpha}^{(k)}$ can be obtained. Note the range of elements of $\boldsymbol{\alpha}^{(k)}$ is (0, 1). Finally, through sequentially performing a soft-thresholding and a multi-layer 1D CNN $\mathcal{N}_2$, the de-aliased 1D image $\mathbf{d}^{(k)}$ outputs. When $k=1$, the initialized network input $\mathbf{x}^{(0)} = \mathcal{F}_{1D}^* \mathcal{U}^* \mathbf{y}$ is the zero-filled 1D image that has strong artifacts.

**2.1.2 Data consistency module**

In this module, each output is ensured to align with the acquired data. The data consistency module is designed mostly same to the second sub-problem of (2) as follows:

$$\mathbf{x}^{(k)} = (\mathcal{F}_{1D}^* \mathcal{U}^* \mathcal{U} \mathcal{F}_{1D} + \lambda^{(k)})^{-1} (\mathcal{F}_{1D}^* \mathcal{U}^* \mathbf{y} + \lambda^{(k)} \mathbf{d}^{(k)}), \tag{5}$$

and the only difference is that we set $\lambda$ as a trainable parameter initialized to 1.

In summary, in the proposed 1D foundation network, a data-versatile image de-aliasing module followed by a data consistency module constitutes a single network phase.

## 2.2 Data-specific k-space enhancement

Beyond the 1D foundation network with data-versatile image de-aliasing, we decide to perform data-specific 2D k-space refinement during the reconstruction stage to enhance the overall network performance. This idea is mainly based on two considerations: 1) Learning from solely synthetic data, the basic network only obtains a versatile deep image prior. Regularizing the solution with another k-space prior learned from target data itself, has potential to generalize the network to realistic data better. 2) The correlation among rows is not exploited in the 1D foundation network and doing some 2D enhancement for compensation should be beneficial.

Thus, we introduce the k-space self-consistency prior, named SPIRiT[10], for enhancement. It refines the output of the learned image module to comply with the 2D linear neighborhood relationship of the target data itself, and is discovered to be effective in details preservation[11].

By introducing the data-specific k-space enhancement module, the overall 2D reconstruction process is:

$$\begin{cases} \mathbf{D}^{(k)} = \sum_{m=1}^{M} \mathcal{P}_m^*(\mathcal{D}(\mathcal{P}_m(\mathbf{X}^{(k-1)}))) = \sum_{m=1}^{M} \mathcal{P}_m^*(\mathbf{d}_m^{(k)}) \\ \mathbf{B}^{(k)} = \mathcal{F}_{2D}^* \mathcal{G}(\mathcal{F}_{2D} \mathbf{D}^{(k)}) \\ \mathbf{X}^{(k)} = \arg\min_{\mathbf{x}} \|\mathbf{Y} - \mathcal{U}\mathcal{F}_{2D}\mathbf{X}\|_2^2 + \lambda \|\mathbf{X} - \mathbf{B}^{(k)}\|_2^2 \\ \qquad = (\mathcal{F}_{2D}^* \mathcal{U}^* \mathcal{U} \mathcal{F}_{2D} + \lambda^{(k)})^{-1}(\mathcal{F}_{2D}^* \mathcal{U}^* \mathbf{Y} + \lambda^{(k)} \mathbf{B}^{(k)}) \end{cases} \quad (6)$$

where $\mathcal{F}_{2D}$ is the 2D FT, $\mathcal{P}_m$ is the operator that extract $m^{th}$ row from each coil data, e.g., $\mathbf{x}_m = \mathcal{P}_m(\mathbf{X})$. $\mathcal{G}$ is the 2D linear convolutional kernel learned from the auto-calibration lines of target data itself.

### 2.3 1D learning meets 2D undersampling

In addition to 1D undersampling, 2D undersampling is also widely adopted in the clinical practice of fast MRI[2]. Although our 1D learning-based method is initially designed for 1D undersampling, it is relatively straightforward to generalize it to handle the case of 2D undersampling, by separately processing rows and columns of the image in the trained de-aliasing modules (Supplementary Fig. 4). Besides, other modules are still consistent with the typical architecture.

In this way, the proposed method can be naturally extended to reconstruct high-quality 2D undersampled MRI images without any re-training.

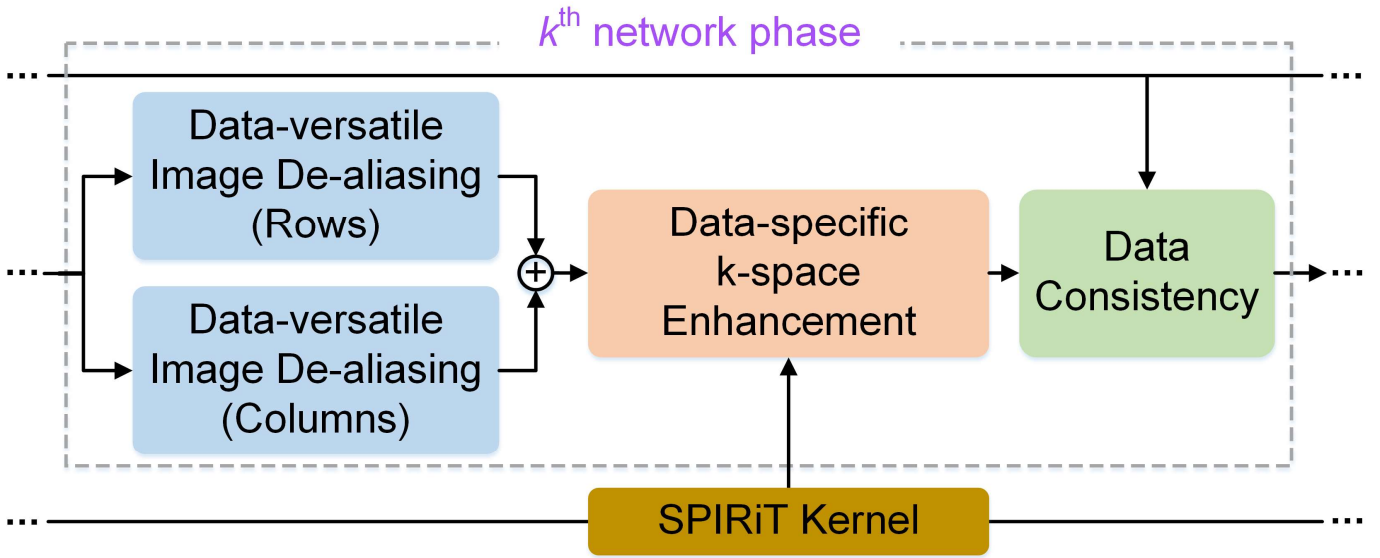

**Supplementary Fig. 4 | The network architecture of the proposed method for 2D undersampling.** By using the trained image de-aliasing module to process the rows and columns of images separately, it is natural to handle the case of 2D undersampling without the need for any re-training.

# Supplementary Note 3. The Justification for Using 1D Synthetic Data

Typically, 2D MRI is accelerated by undersampling the 1D phase encoding (PE) of a 2D k-space, while another frequency encoding (FE) is fully sampled[2,12,13]. It means that each 1D data shares the same 1D undersampling mask. Therefore, for a 2D MRI reconstruction problem, we could separate it into many 1D basic problems to reduce the dimensionality/diversity of the solution space[14,15]. Existing work shows that compared to the common 2D learning, the 1D learning can make the deep network easier to train and have better robustness against mismatches in training and target data[16].

We further added a comparison experiment on networks trained with 1D and 2D synthetic data, and we named them as PISF (1D) and PISF (2D), respectively. For the latter version, 2D synthetic data was generated thus we only needed to change the size of convolution filters from 1D to 2D, while keeping other network architectures unchanged.

Supplementary Fig. 5 shows that the 1D version of PISF outperforms its 2D version visually and quantitatively in reconstructions of all three anatomies, indicating the good ability of artifacts suppression and details preservation. While the 2D version yields obvious artifacts in brain and cardiac reconstructions. 2D learning implies more trainable network parameters, which are nearly three times than those of the 1D learning. Then, insufficient training data may not provide faithful results and our 1D learning can overcome this limitation.

Besides, as aforementioned, although 1D synthetic data cannot capture higher dimensional correlations in the training stage, our PISF integrates a data-specific 2D k-space enhancement obtained from target data itself (self-consistency refinement[10,11]) in the reconstruction stage, to enhance the correlation between rows and columns of k-space.

Thus, to develop a generalizable deep learning method, we choose to use 1D data synthesis instead of 2D data in this work.

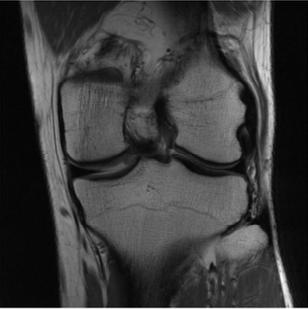

**Supplementary Fig. 5 | Comparison results on networks trained with 1D and 2D synthetic data.** Here, they include 3 anatomies under 2 undersampling scenarios (Data I, II, IX). Note: The mean values of PSNR are computed over all tested cases.

# Supplementary Note 4. Network Physical Behaviors and Integrations

The proposed PISF not only encourages physical behaviors, but also integrates explicit physical operators into the network architecture. It belongs to physics-informed or physics-driven MRI reconstruction[9,17-21].

We propose an unrolled network that follows an iteration process of a reconstruction model with image and k-space priors, rather than a simple end-to-end network architecture. Similar unrolled network strategies have been observed to provides stable reconstruction[9,22]. Intermediate reconstructions (Supplementary Fig. 6) show that, like model-based methods, PISF gradually improves images both visually and quantitatively with the increase of network phases. This visualization provides a good interpretation to understand the network behavior.

The proposed network also integrates three explicit physical operators, including data consistency [2,23], soft-thresholding[16,24], and self-consistency refinement[10,11]. Specifically, 1) Data consistency can ensure that reconstructed k-space are aligned to measured data, to improve the network reliability. 2) Soft-thresholding is always used to enforce the sparsity in signal processing and can remove image artifacts and noises efficiently through threshold adjustments. 3) Self-consistency refinement is integrated to enhance the correlation between rows and columns of k-space, which is beneficial for image details preservation.

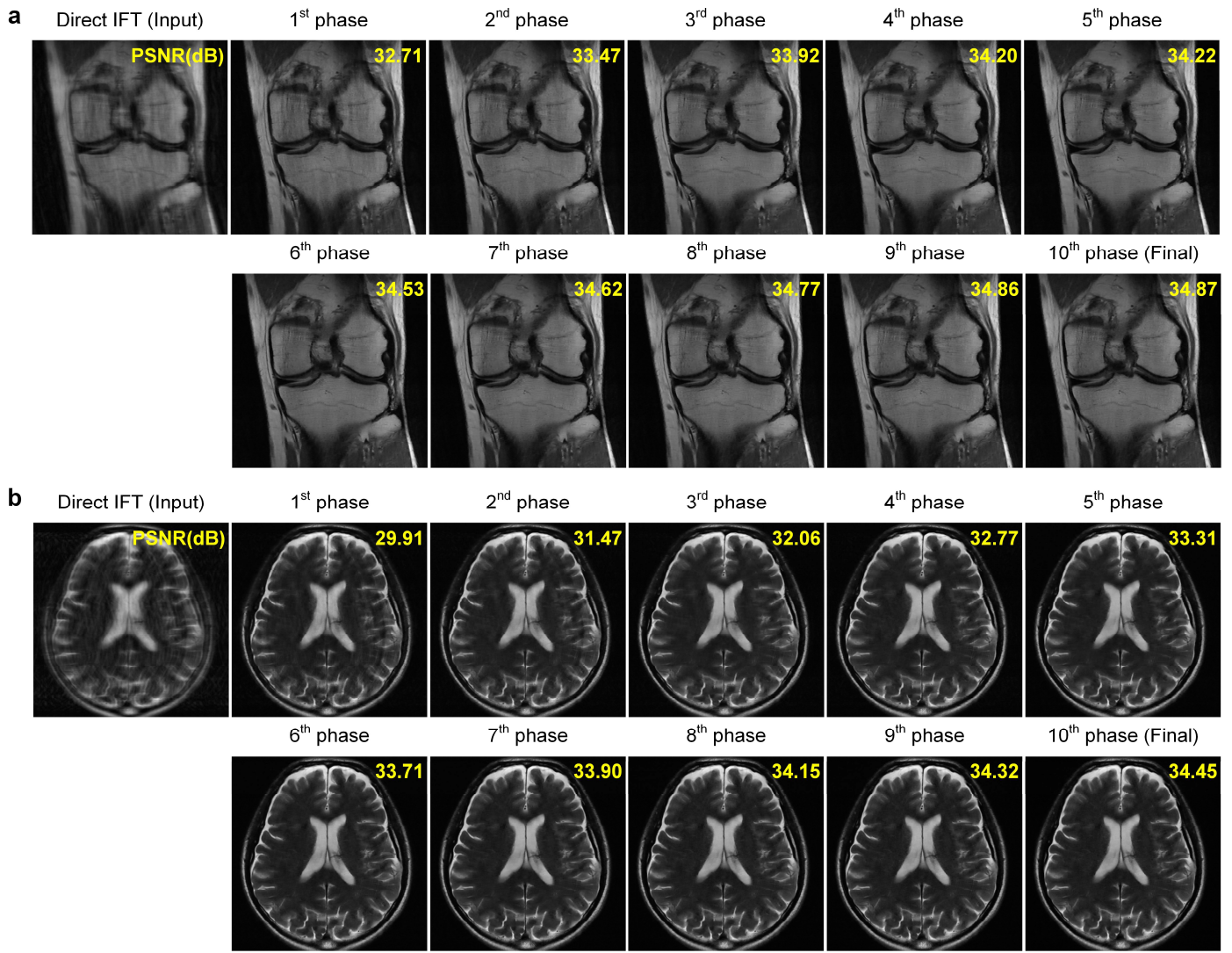

**Supplementary Fig. 6 | Visualization of the reconstructed images at each network phase of PISF. a**, PDW knee reconstruction results (Data I). **b**, T2W brain reconstruction results (Data II). Note: The 1D Cartesian undersampling pattern with AF=4 is used.

# Supplementary Note 5. Data Synthesis with Varied Number of Realistic Cases

We further compared our PISF under different number of realistic cases which are used for coil sensitivity maps estimation in data synthesis. Supplementary Figs. 7(a)-(b) show that, if no realistic case is used, the reconstruction performance is worst. This is because we need to simulate the coil sensitivity maps by using the Biot-Savart law[25] for data synthesis, which is too ideal for realistic scenarios. Interestingly, when only 1 realistic case is used, the PSNR can be significantly improved, indicating the necessity of realistic sensitivity information. Then, with the increase of this number, the improvements of PISF still exist but gradually become much smaller. Thus, to balance the reconstruction performance and realistic data usage, we choose to use 6 realistic cases for data synthesis in our work.

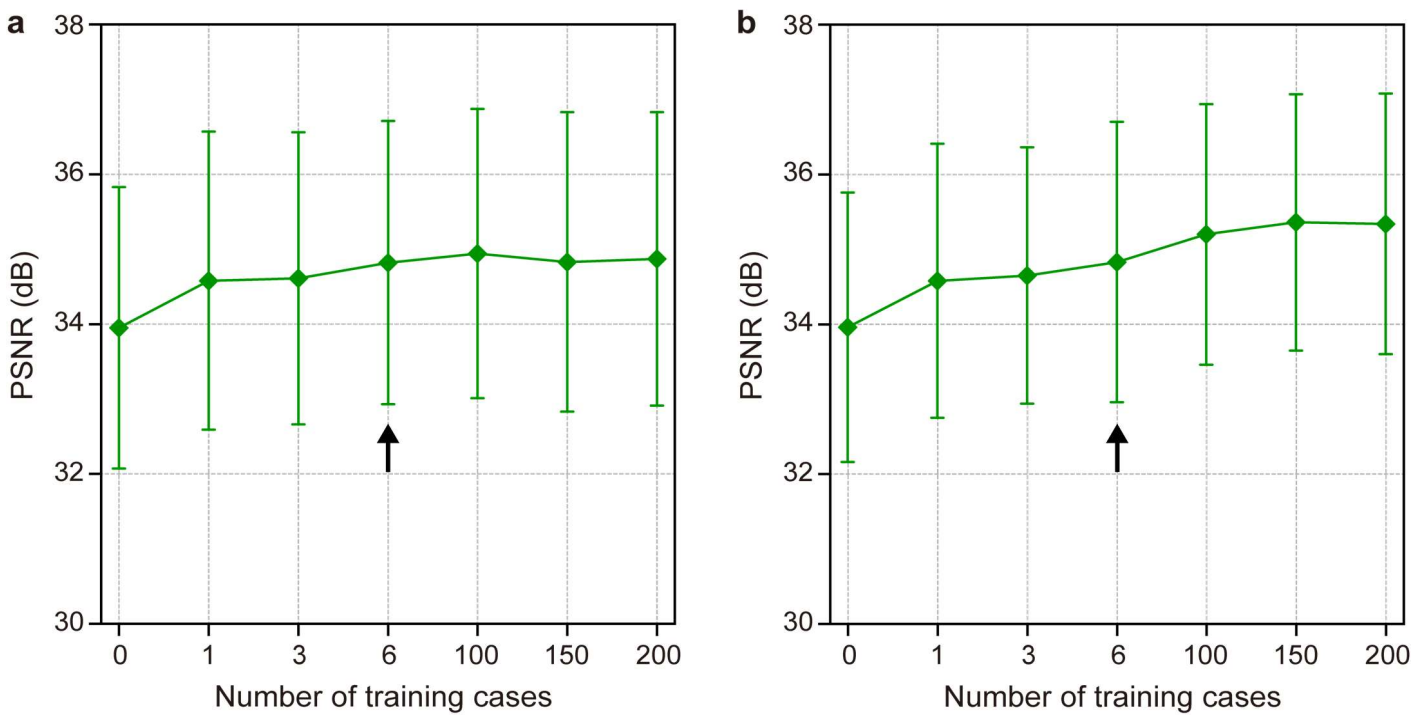

**Supplementary Fig. 7 | PSNRs of knee and brain reconstructions using different number of realistic training cases in PISF. a**, PDW knee reconstruction results (Data I). **b**, T2W brain reconstruction results (Data II). Note: PISF only estimates coil sensitivity maps from the autocalibration signals of realistic data for data synthesis, which means no fully sampled MRI data is needed. The black arrow denotes the chosen setting in the paper. The 1D Cartesian undersampling pattern with AF=4 is used. The mean values and standard deviations of PSNR are computed over all tested cases.

# Supplementary Note 6. More Results of Multi-vendor Multi-center Imaging

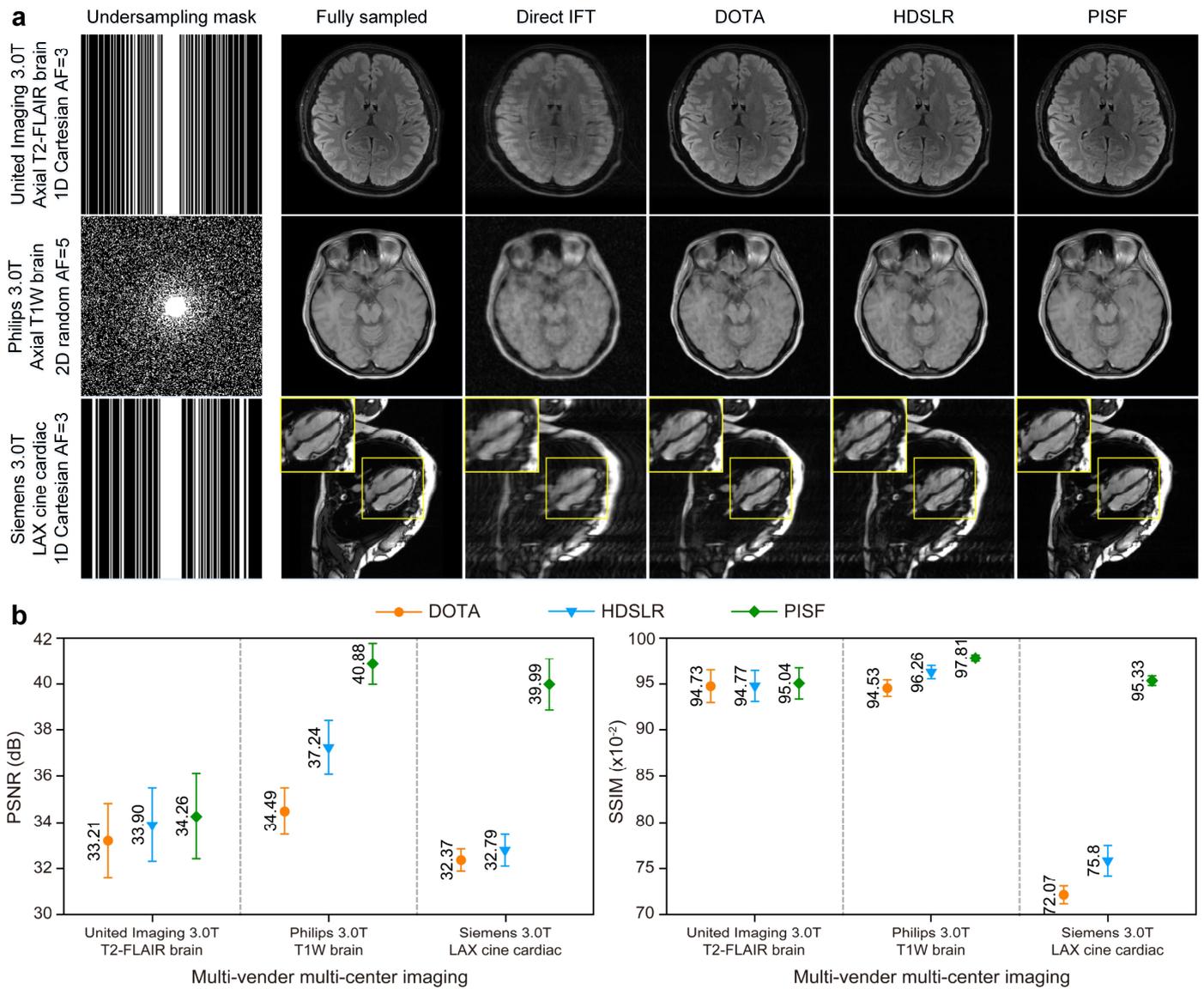

**Supplementary Fig. 8 | More results of multi-vendor multi-center imaging. a**, multi-scenario reconstruction results using different methods. Here, they include 2 anatomies and 3 contrasts across 3 vendors and centers, under 2 undersampling scenarios (Data VI, VII, XIV). **b**, quantitative comparisons of reconstructions are shown, including PSNR and SSIM. Note: The mean values and standard deviations of PSNR and SSIM are computed over all tested cases, respectively.

# Supplementary Note 7. Matched/Mismatched Undersampling Reconstruction

Here, we provide the typical reconstruction results under matched and mismatched undersampling settings. In these experiments, the training datasets of DOTA and HDSLR are realistic fastMRI dataset (including knee and brain), while our PISF uses synthetic data for training. We conduct two types of experiments, called "***Matched undersampling settings and anatomical structures***" and "***Matched undersampling settings but mismatched anatomical structures***", to explore the impact of undersampling settings and anatomical structures of training datasets on reconstruction.

*Matched undersampling settings and anatomical structures:* Supplementary Figs. 9 and 10 show that, for two brain reconstructions, matched undersampling settings can improve the image quality of all three methods, and the proposed PISF consistently provides superior results visually and quantitatively.

*Matched undersampling settings but mismatched anatomical structures:* But for the cardiac reconstruction (Supplementary Fig. 11), only PISF benefits from the matched undersampling setting, while the other two compared methods are even worse. This may be due to the anatomical mismatch that still exists in this experiment, and it indicates that the proposed PISF can effectively alleviate the performance degradation in this scenario.

These results also imply that the matching of anatomical structures in training and target data is crucial for high-performance reconstruction, while the impact of undersampling settings is relatively low.

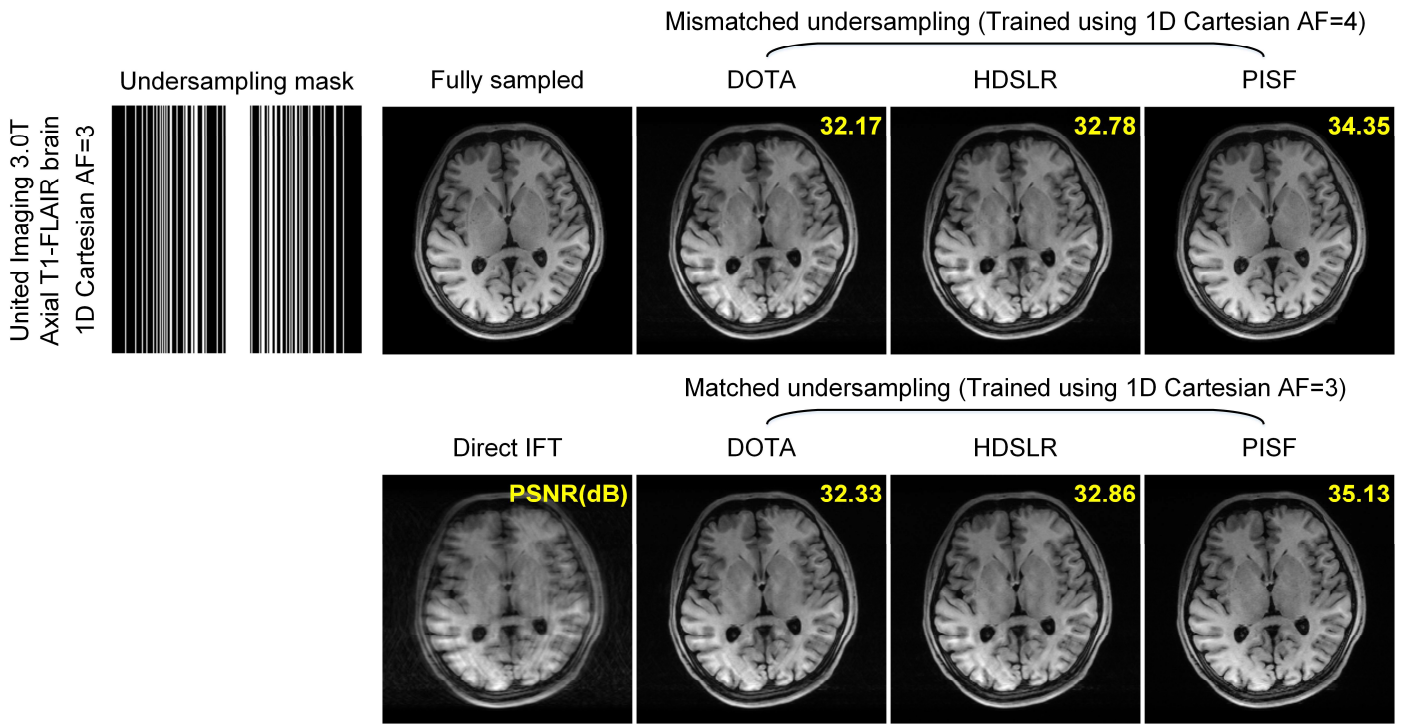

**Supplementary Fig. 9 | Comparison results under mismatched and matched undersampling settings (Data V).** Note: In mismatched undersampling settings, all three networks are trained using the 1D Cartesian undersampling pattern with AF=4, while in matched settings, they are trained using undersampling settings of target data. The training datasets of DOTA and HDSLR are realistic fastMRI dataset, while our PISF uses synthetic data for training. The mean values of PSNR are computed over all tested cases.

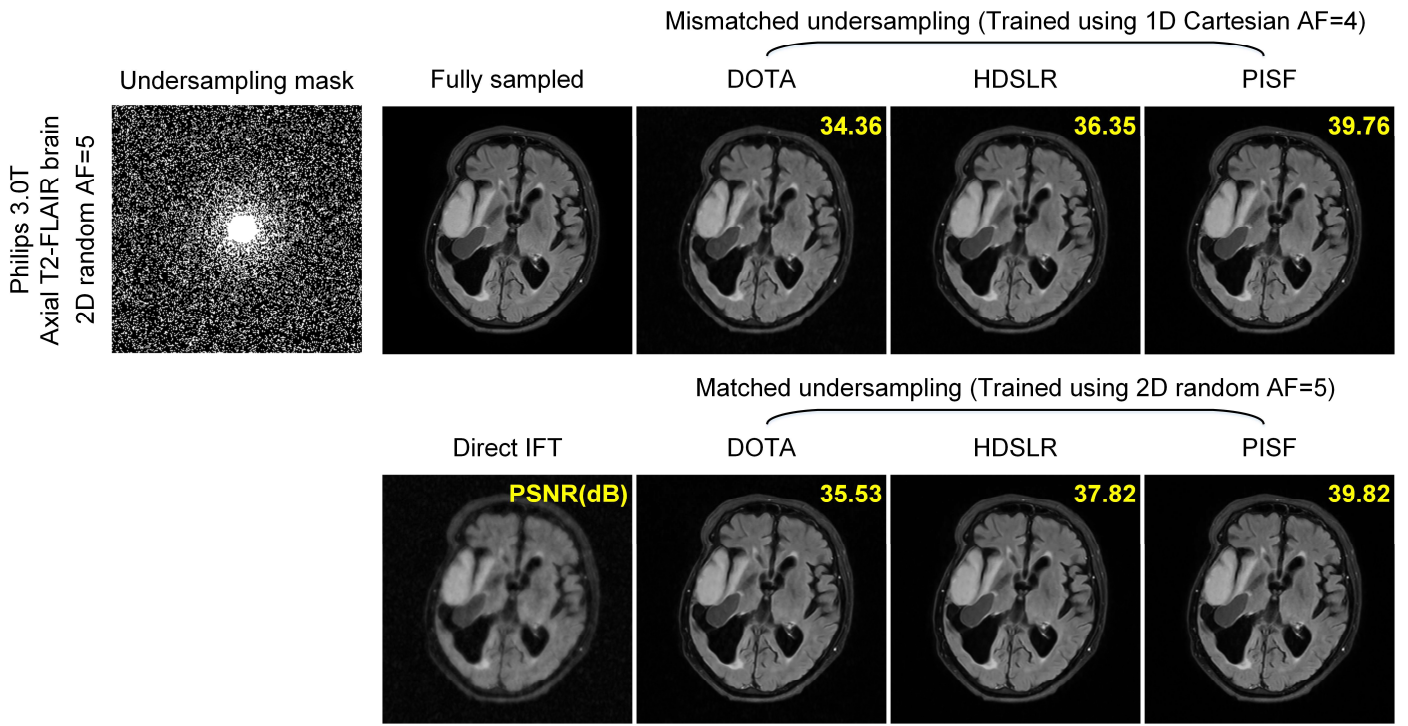

**Supplementary Fig. 10 | Comparison results under mismatched and matched undersampling settings (Data XII).** Note: In mismatched undersampling settings, all three networks are trained using the 1D Cartesian undersampling pattern with AF=4, while in matched settings, they are trained using undersampling settings of target data. The training datasets of DOTA and HDSLR are realistic fastMRI dataset, while our PISF uses synthetic data for training. The mean values of PSNR are computed over all tested cases.

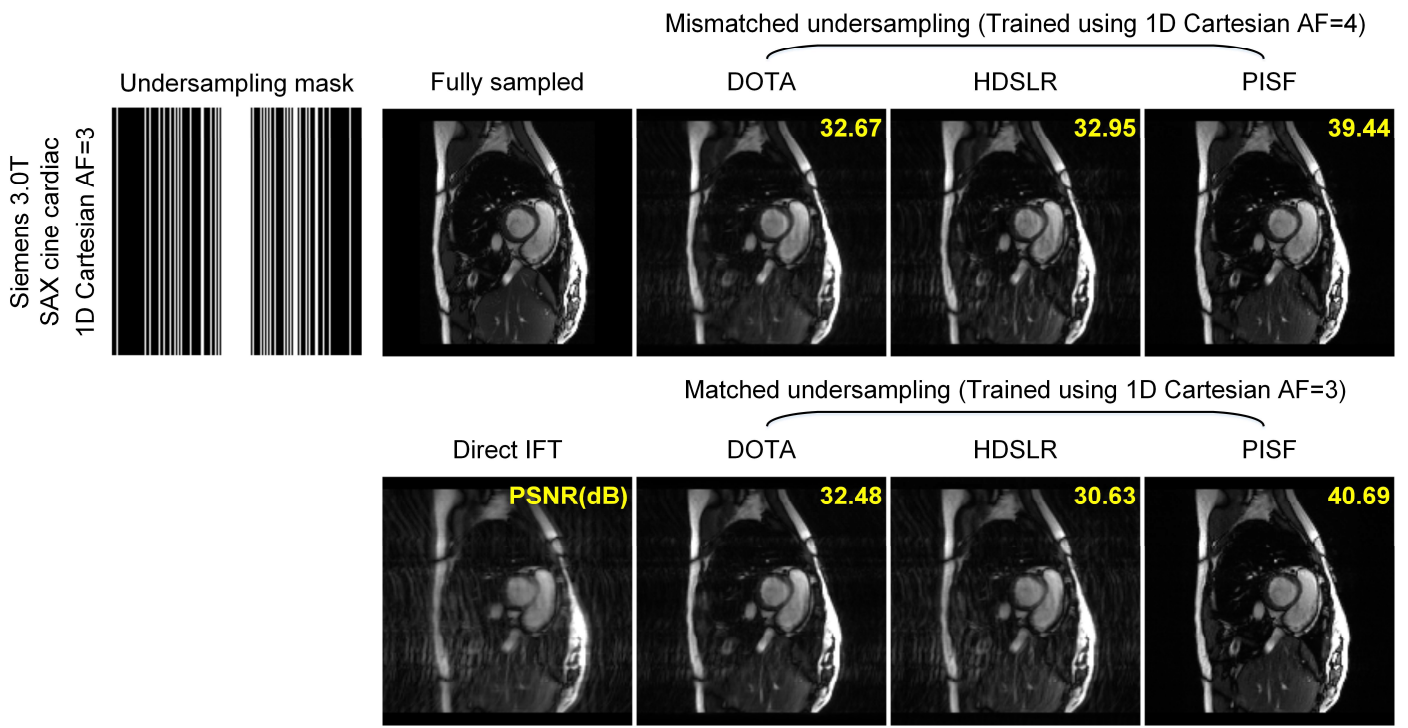

**Supplementary Fig. 11 | Comparison results under mismatched and matched undersampling settings (Data IX).** Note: In mismatched undersampling settings, all three networks are trained using the 1D Cartesian undersampling pattern with AF=4, while in matched settings, they are trained using undersampling settings of target data. The training datasets of DOTA and HDSLR are realistic fastMRI dataset, while our PISF uses synthetic data for training. The mean values of PSNR are computed over all tested cases.

# Supplementary Note 8. Scores of All Individual Readers

**Supplementary Table 1 | Scores of all individual readers of axial T2W brain data with non-specific white matter lesion (Data XI).**

| Reader | Criterion | DOTA | HDSLR | PISF |
|---|---|---|---|---|
| #1 | SNR | 4.35±0.13 | **4.36±0.14** | 4.34±0.13 |
| | Artifacts suppression | 4.34±0.14 | 4.33±0.15 | **4.35±0.15** |
| | Overall image quality | 4.35±0.14 | 4.35±0.13 | **4.36±0.12** |
| #2 | SNR | 4.39±0.17 | 4.41±0.16 | **4.44±0.14** |
| | Artifacts suppression | 4.38±0.17 | 4.41±0.17 | **4.43±0.14** |
| | Overall image quality | 4.39±0.21 | 4.43±0.15 | **4.45±0.15** |
| #3 | SNR | 4.16±0.82 | 4.20±0.80 | **4.21±0.81** |
| | Artifacts suppression | 4.15±0.83 | 4.20±0.80 | **4.21±0.82** |
| | Overall image quality | 4.16±0.82 | 4.20±0.80 | **4.22±0.81** |
| #4 | SNR | 3.78±0.62 | 4.00±0.70 | **4.17±0.62** |
| | Artifacts suppression | 3.78±0.62 | 4.05±0.67 | **4.14±0.65** |
| | Overall image quality | 3.75±0.63 | 4.07±0.68 | **4.10±0.68** |
| #5 | SNR | 4.47±0.15 | 4.53±0.14 | **4.65±0.13** |
| | Artifacts suppression | 4.54±0.14 | 4.58±0.12 | **4.64±0.10** |
| | Overall image quality | 4.47±0.17 | 4.62±0.16 | **4.70±0.12** |

Note: 5 neuro readers (4 radiologists with 6/13/20/30 years' experience and 1 neurosurgeon with 9 years' experience) independently evaluate brain reconstructed images from a diagnostic perspective. The mean values and standard deviations are computed over all tested patients, respectively. The highest scores are bold faced.

**Supplementary Table 2 | Scores of all individual readers of axial T2-FLAIR brain data with tumor (Data XII).**

| Reader | Criterion | DOTA | HDSLR | PISF |
|---|---|---|---|---|
| #1 | SNR | 4.12±0.24 | 4.12±0.21 | **4.30±0.22** |
| | Artifacts suppression | 4.12±0.22 | 4.13±0.21 | **4.28±0.21** |
| | Overall image quality | 4.11±0.21 | 4.14±0.21 | **4.29±0.25** |
| #2 | SNR | 4.13±0.10 | 4.42±0.12 | **4.46±0.10** |
| | Artifacts suppression | 4.25±0.14 | 4.41±0.12 | **4.46±0.10** |
| | Overall image quality | 4.23±0.14 | 4.45±0.12 | **4.48±0.10** |
| #3 | SNR | 3.19±0.45 | 3.81±0.40 | **4.12±0.49** |
| | Artifacts suppression | 3.23±0.52 | 3.87±0.46 | **4.14±0.50** |
| | Overall image quality | 3.22±0.53 | 3.87±0.45 | **4.15±0.47** |
| #4 | SNR | 3.07±0.42 | 3.04±0.28 | **4.07±0.29** |
| | Artifacts suppression | 3.09±0.43 | 3.04±0.32 | **4.07±0.28** |
| | Overall image quality | 3.08±0.40 | 3.05±0.26 | **4.06±0.30** |
| #5 | SNR | 4.50±0.17 | 4.53±0.14 | **4.65±0.13** |
| | Artifacts suppression | 4.48±0.19 | 4.58±0.12 | **4.64±0.10** |
| | Overall image quality | 4.48±0.21 | 4.62±0.16 | **4.70±0.12** |

Note: 5 neuro readers (4 radiologists with 6/13/20/30 years' experience and 1 neurosurgeon with 9 years' experience) independently evaluate brain reconstructed images from a diagnostic perspective. The mean values and standard deviations are computed over all tested patients, respectively. The highest scores are bold faced.

**Supplementary Table 3 | Scores of all individual readers of SAX cine cardiac data with myocardial hypertrophy (Data XIII).**

| Reader | Criterion | DOTA | HDSLR | PISF |
|---|---|---|---|---|
| #1 | SNR | 3.10±0.46 | 2.65±0.65 | **4.07±0.60** |
| | Artifacts suppression | 3.00±0.52 | 2.47±0.74 | **3.95±0.60** |
| | Overall image quality | 3.33±0.52 | 2.83±0.66 | **4.27±0.55** |
| #2 | SNR | 2.99±0.27 | **3.46±0.32** | 3.19±0.19 |
| | Artifacts suppression | 2.96±0.23 | 2.16±0.40 | **3.89±0.36** |
| | Overall image quality | 2.96±0.18 | 2.80±0.19 | **3.55±0.21** |
| #3 | SNR | 2.90±0.58 | 1.49±0.43 | **4.76±0.19** |
| | Artifacts suppression | 2.81±0.58 | 1.44±0.42 | **4.73±0.26** |
| | Overall image quality | 2.96±0.59 | 1.52±0.42 | **4.79±0.20** |
| #4 | SNR | 4.29±0.24 | 3.83±0.44 | **4.76±0.22** |
| | Artifacts suppression | 3.73±0.51 | 3.03±0.79 | **4.67±0.36** |
| | Overall image quality | 4.00±0.36 | 3.42±0.67 | **4.71±0.27** |
| #5 | SNR | 3.91±0.40 | 3.78±0.28 | **4.34±0.31** |
| | Artifacts suppression | 3.87±0.39 | 3.73±0.26 | **4.34±0.32** |
| | Overall image quality | 4.34±0.24 | 4.23±0.24 | **4.60±0.15** |

Note: 5 cardiac readers (3 radiologists with 3/8/12 years' experience and 2 cardiologists with 12/13 years' experience) independently evaluate cardiac reconstructed images from a diagnostic perspective. The mean values and standard deviations are computed over all tested patients, respectively. The highest scores are bold faced.

**Supplementary Table 4 | Scores of all individual readers of LAX cine cardiac data with myocardial hypertrophy (Data XIV).**

| Reader | Criterion | DOTA | HDSLR | PISF |
|---|---|---|---|---|
| #1 | SNR | 2.49±1.07 | 2.18±1.02 | **3.23±1.11** |
| | Artifacts suppression | 2.50±1.10 | 2.22±1.03 | **3.33±1.08** |
| | Overall image quality | 2.76±1.02 | 2.42±1.07 | **3.63±1.02** |
| #2 | SNR | 2.73±0.50 | 3.24±0.48 | **3.35±0.38** |
| | Artifacts suppression | 2.87±0.52 | 2.16±0.52 | **3.68±0.47** |
| | Overall image quality | 2.79±0.35 | 2.69±0.39 | **3.44±0.53** |
| #3 | SNR | 3.00±0.98 | 1.50±0.32 | **4.26±0.59** |
| | Artifacts suppression | 2.73±0.98 | 1.29±0.21 | **4.02±0.72** |
| | Overall image quality | 2.99±0.92 | 1.52±0.29 | **4.23±0.61** |
| #4 | SNR | 3.78±0.37 | 3.65±0.47 | **4.39±0.31** |
| | Artifacts suppression | 3.28±0.86 | 2.67±1.02 | **4.28±0.54** |
| | Overall image quality | 3.34±0.69 | 2.94±0.89 | **4.29±0.44** |
| #5 | SNR | 3.43±0.51 | 3.38±0.52 | **4.09±0.42** |
| | Artifacts suppression | 3.43±0.50 | 3.33±0.51 | **4.11±0.44** |
| | Overall image quality | 3.73±0.59 | 3.63±0.55 | **4.31±0.34** |

Note: 5 cardiac readers (3 radiologists with 3/8/12 years' experience and 2 cardiologists with 12/13 years' experience) independently evaluate cardiac reconstructed images from a diagnostic perspective. The mean values and standard deviations are computed over all tested patients, respectively. The highest scores are bold faced.

# Supplementary Note 9. Taking PISF as A Pre-trained Model

Although our proposed PISF has shown its robustness in multi-scenario MRI reconstructions, here, we extend it as a pre-trained model to further improve its performance using transfer learning.

Supplementary Fig. 12 shows that taking PISF as a pre-trained model, only one case of target data is good enough for transfer learning to provide improved reconstruction both visually and quantitatively, thus we call this extension as PISF+1. Besides, Supplementary Fig. 12(b) demonstrates that, at a higher acceleration (AF=5), compared to the pre-trained PISF, the visual and quantitative gains brought by PISF+1 are more significant, which can further remove residual artifacts and improve PSNR by 1.09dB.

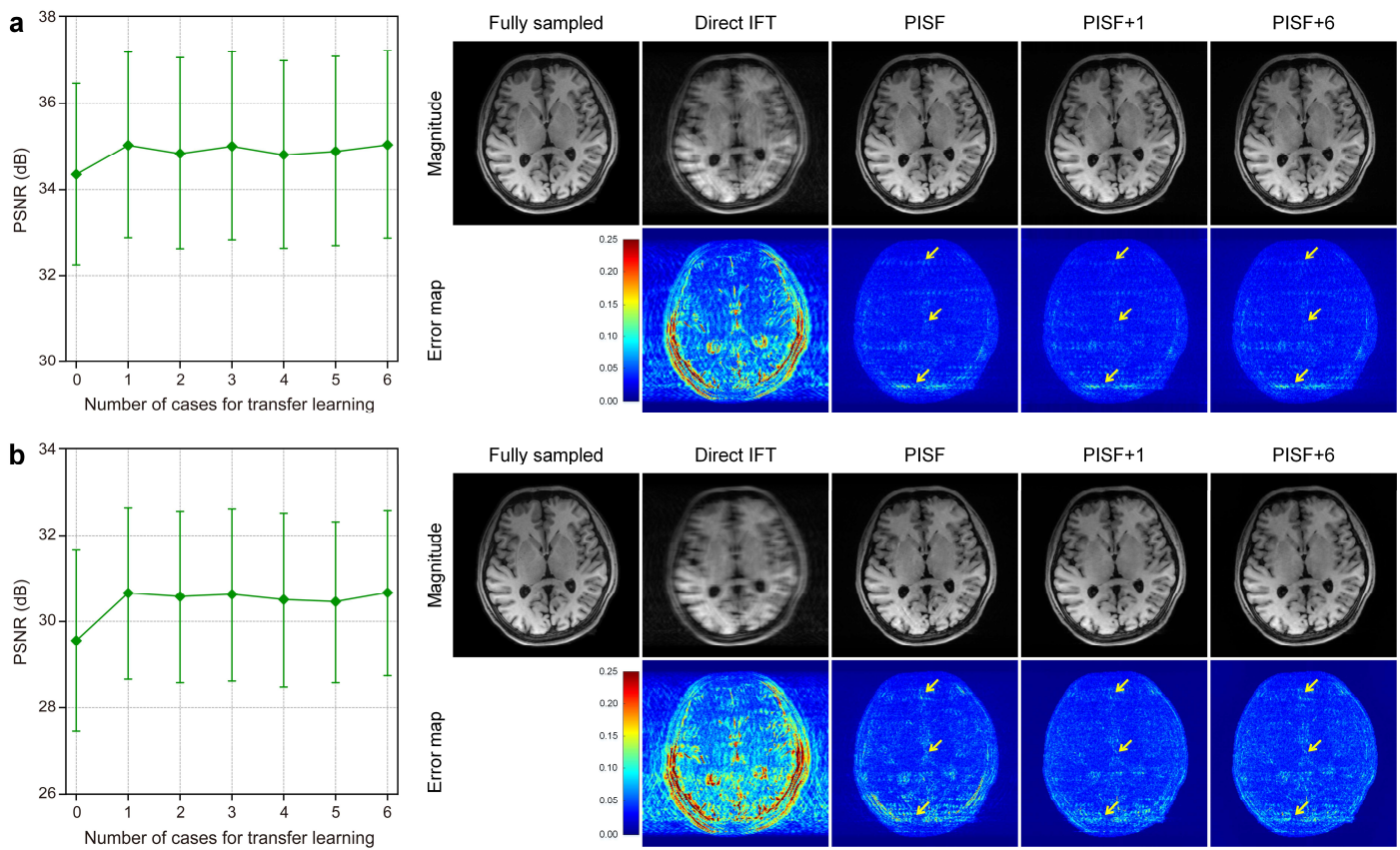

**Supplementary Fig. 12 | Brain reconstruction results using different number of target cases for transfer learning under different undersampling settings (Data V). a**, results under the 1D Cartesian undersampling pattern with AF=3. **b**, results under the 1D Cartesian undersampling pattern with AF=5. Note: PISF is used as a pre-trained model. "PISF+1" and "PISF+6" represent using one and six cases of target realistic brain data for transfer learning, respectively. The improvements are marked with yellow arrows. The mean values and standard deviations of PSNR are computed over all tested cases.

# Supplementary Note 10. Insights on Synthetic Data for Existing Methods

For existing deep learning methods trained using realistic data (take HDSLR[26] trained using fastMRI dataset as an example in Supplementary Fig. 13), the reconstruction performance is poor when encountering unseen data (i.e., cardiac images are not included in the used fastMRI dataset). At this time, our synthetic data can become a powerful supplement to enhance the generalizability of HDSLR. However, our proposed PISF still outperforms the synthetic-data-aided HDSLR, implying the superiority of our network architecture.

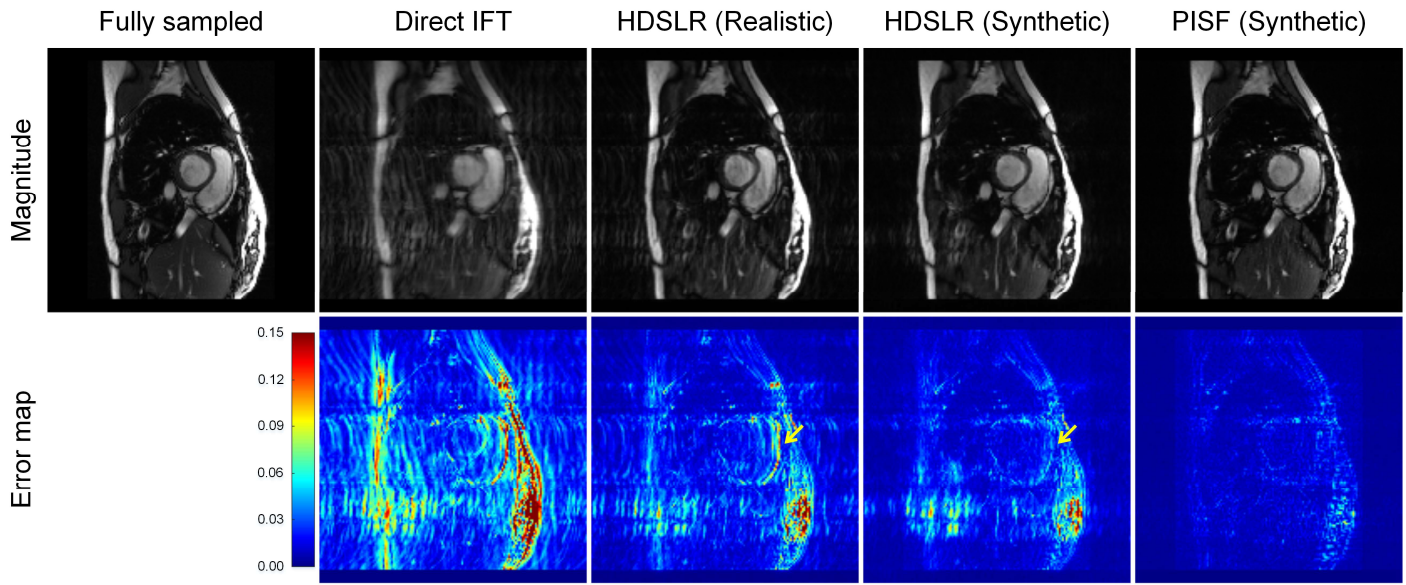

**Supplementary Fig. 13 | Cardiac reconstruction results of different methods under the 1D Cartesian undersampling pattern with AF=3 (Data IX).** Note: The obvious improvements are marked with yellow arrows after using synthetic data for supplementary training.

# Supplementary Note 11. Limitation on High Acceleration Factors

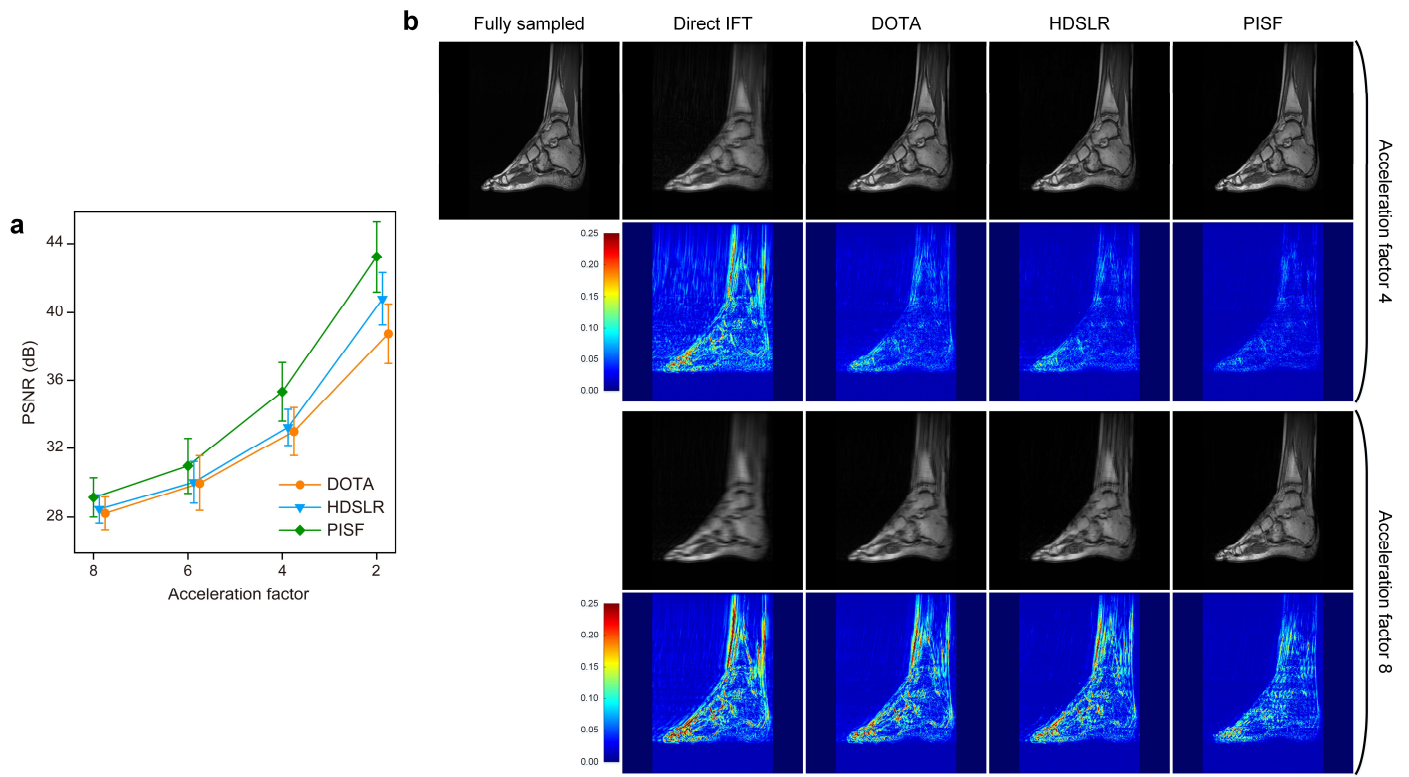

**Supplementary Fig. 14 | Limitation on high acceleration factors. a**, PDW ankle (Data III) reconstruction results of different methods using the 1D Cartesian undersampling pattern with different acceleration factors (AFs). **b**, reconstructed images and error maps of different methods under different AFs. Note: The mean values and standard deviations of PSNR are computed over all tested cases. Even though, our PISF demonstrates more robust results than compared methods, it still cannot provide satisfactory reconstructions under extremely high acceleration (AF=8).